\documentclass[a4paper,11pt]{article}

\usepackage[english]{babel}
\usepackage[latin1]{inputenc}
\usepackage{indentfirst}
\usepackage{fancyhdr}
\usepackage{amsmath, amssymb, mathrsfs, dsfont}
\usepackage{amsthm}
\usepackage{empheq}
\usepackage{latexsym}
\usepackage{graphicx}
\usepackage{float, placeins}
\usepackage{wrapfig}
\usepackage[textfont=sl,font=small,format=plain,labelfont={sf,bf}]{caption}
\usepackage{subfig}
\usepackage{mathtools}
\usepackage{comment}
\usepackage{enumitem}
\usepackage{rotating}
\usepackage{color}

\addto\captionsenglish{}

\newcommand{\N}{\mathbb N}
\newcommand{\R}{\mathbb R}

\newcommand{\Q}{\mathbb Q}
\newcommand{\C}{\mathbb C}
\newcommand{\HH}{\mathbb H}
\newcommand{\E}{\mathbb E}
\newcommand{\Prob}{\mathbb P}
\newcommand{\1}{\mathds 1}

\newcommand{\F}{\mathcal F}
\newcommand{\dd}{\mathrm d}

\DeclarePairedDelimiter{\bracket}{\langle}{\rangle}

\newcommand{\tsum}{\textstyle\sum}

\newcommand{\tprod}{\textstyle\prod}
\newcommand{\DD}{\mathscr D}
\newcommand{\MM}{\mathscr M}
\newcommand{\RR}{\mathcal R}
\newcommand{\EE}{\mathcal E}
\newcommand{\T}{\mathcal T}
\newcommand{\G}{\mathscr G}
\newcommand{\TT}{\mathscr T}
\newcommand{\w}{\mathbf w}
\newcommand{\x}{\mathbf x}
\newcommand{\Card}{\textit{Card}}
\newcommand{\vn}{o}
\newcommand{\Po}{\overline{P}}
\newcommand{\rhoo}{\overline{\rho}}

\theoremstyle{plain}
\newtheorem{thm}{Theorem}
\newtheorem{lem}{Lemma}
\newtheorem{prop}{Proposition}
\newtheorem{cor}{Corollary}

\theoremstyle{definition}
\newtheorem{df}{Definition}
\newtheorem{rk}{Remark}
\newtheorem{ex}{Example}

\title{Solution of the monomer-dimer model on locally tree-like graphs. Rigorous results.}
\author{Diego Alberici and Pierluigi Contucci\footnote{Universit\`a di Bologna, piazza di Porta San Donato 5, 40127 Bologna, Italy. E-mail addresses: diego.alberici2@unibo.it, pierluigi.contucci@unibo.it}}

\begin{document}

\maketitle

\begin{abstract}
We consider the monomer-dimer model on sequences of random graphs locally convergent to trees.
We prove that the monomer density converges almost surely, in the thermodynamic limit, to an analytic function  
of the monomer activity. We characterise this limit as the expectation of the solution of a fixed point 
distributional equation and we give an explicit expression for the limiting pressure per particle.
\end{abstract}

\section{Introduction}
Each way to fully cover the vertices of a finite graph $G$ by non-overlapping monomers (molecules which occupy a single vertex) and dimers (molecules which occupy two adjacent vertices) is called a monomer-dimer configuration.
Associating to each of those configurations a probability proportional to the product of a factor $w'>0$ for each dimer and a factor $x'>0$ for each monomer defines a monomer-dimer model on the graph.
It is easily seen that 
the monomer-dimer probability measure depends only on the factor $x=x'/\sqrt{w'}\,$.
What one is mainly interested in are the monomer (and dimer) densities, i.e. the average number of monomers (dimers) per site.

Monomer-dimer models were introduced in the last century in the physics literature to study the statistical mechanics problem of diatomic oxygen adsorption on tungsten \cite{R} and similar phenomena (see \cite{HL} and references therein).
Important rigorous results were obtained by Heilmann and Lieb in \cite{HLprl,HL}, where in particular the absence of phase transition for the pressure as a function of $x$ (and of the monomer density too) was proved for all positive $x$.
Furthermore in \cite{HL} exact solutions were given for specific topologies like the one-dimensional (with free and periodic boundary conditions), the complete graph and the Bethe lattice.
Previously exact solutions on two-dimensional lattices where found by Kasteleyn, Fisher and Temperley \cite{K,F,TF} for the pure dimer problem, i.e. the problem of counting configurations with no monomers. 

In this paper we study the statistical mechanics properties of monomer-dimer systems on locally tree-like random graphs 
computing their monomer density and their pressure (see also \cite{AD}) in the thermodynamic limit for all positive $x$.
The class of diluted graphs that we cover is the same for which the exact solution of the ferromagnetic Ising model was recently found by Dembo and Montanari \cite{DM} using the local weak convergence strategy developed in \cite{A};
precisely we consider random graphs $(G_n)_{n\in\N}$ locally convergent to a unimodular Galton-Watson tree $\T(P,\rho)$ and with finite second moment of the asymptotic degree distribution $P$.
A remarkable example is the Erd\H{o}s-R\'enyi graph, i.e. the complete graph randomly diluted with i.i.d. Bernoulli edges and average degree $c$.

In the Erd\H{o}s-R\'enyi case our main result is the proof that the monomer density $\varepsilon(x)$ and the pressure 
per particle $p(x)$ exist in the thermodynamic limit and are analytic functions. $\varepsilon(x)$ turns out to be the 
expected value of a random variable $Y(x)$ whose distribution is defined as the only fixed point supported in 
$[0,1]$ of the distributional equation
\begin{equation}\label{ce}
Y \,\overset{\cal D}{=}\, \frac{x^2}{x^2+\sum_{i=1}^K Y_i} \;,
\end{equation}
where the $(Y_i)_{i\in\N}$ are i.i.d.$\!$ copies of $Y$, $K$ is Poisson($c$)-distributed and independent of $(Y_i)_{i\in\N}$.
$p(x)$ is shown to be
\begin{equation}\label{pr} 
-\; \E\big[\log\frac{Y(x)}{x}\,\big] \,-\, \frac{c}{2}\ \E\big[\log\big(1+\frac{Y_1(x)}{x}\;\frac{Y_2(x)}{x}\big)\big] \;. 
\end{equation}
A side-result, yet a crucial one, of our analysis is that the solution $Y(x)$ is reached monotonically in the number of iterations of  equation (\ref{ce}). 
More precisely, starting from $Y_i\equiv 1$, the even iterations decrease monotonically, the odd ones increase monotonically (see Fig.2), their difference shrinks to zero and their common limit is an analytic function of $x$.

Our results are built on the Heilmann-Lieb recursion relation for the partition function $Z_G(x)$ of a monomer-dimer system \cite{HL}.
Given a finite graph $G$, a root vertex $o$ and its neighbours $v$, it holds:
\begin{equation}\label{ipf}
Z_G(x) \,=\, x\,Z_{G-o}(x) + \sum_{v\sim o} Z_{G-o-v}(x)\, .
\end{equation}
In \cite{BLS} it is shown how to rewrite the identity (\ref{ipf}) in terms of the probability $\RR_x(G,o)$ of having a monomer 
in $o$.
We use this form to deduce the distributional identity (\ref{ce}) for $Y(x):=\lim_{r\to\infty}\RR_x(\T(r),o)$, where $\T(r)$ is a random tree with root $o$, $r$ generations and i.i.d.$\!$ Poisson($c$) offspring sizes.
Our results rely on a correlation inequality method which we prove in several forms (see Lemma \ref{lemma: localisation} and the Appendix) and which permits a local study of the monomer-dimer system and is also at the origin of the monotonicity property described before. Analytic continuation techniques are used in order to extend results from ``large'' $x$ to all positive $x$. 

Our results extend those of Bordenave, Lelarge, Salez \cite{BLS} which are valid for graphs with bounded degree, and are generalised to arbitrary degree only for $x\rightarrow 0$ (this is called the maximum matching problem and it is not treated in this paper, see instead \cite{KS} for its first solution in the Erd\H{o}s-R\'enyi case and \cite{S,Le} for other generalisations).
A complete theoretical physics picture of the monomer-dimer model (matching problem) on sparse random graphs was given by Zdeborov\'a and M\'ezard in \cite{ZM}, where several quantities were computed including the pressure of the model,
using the so called replica-symmetric version of the cavity method.
Then Bayati and Nair \cite{BN} obtained rigorous results for a class of graphs satisfying a quite restrictive large girth condition.
Salez \cite{S}, using the language of cavity method, made a rigorous study for locally tree-like graphs that partially overlap with the one presented here and deals also with non hard-core dimer interactions (\textit{b}-matching).

The paper is organised as follows.
Section 2 introduces the definitions and the basic properties of the monomer-dimer models, including their well-posedness with stability bounds for the pressure, the main recursion relation for $\RR_x(G,o)$, the analyticity property of its solutions, and the correlation inequalities for locally tree-like graphs at even and odd tree depth.
Section 3 studies the model on trees, in particular the solution on a Galton-Watson tree is found in Theorem \ref{th: Galton-Watson} and its corollaries.
Section 4 presents the general solution on locally tree-like graphs in Theorem \ref{main theorem}, its corollaries and Theorem \ref{thm: pressure limit}.
Section 5 displays lower and upper bounds for the monomer density in the Erd\H{o}s-R\'enyi case, obtained by iterating the recursion relation (\ref{ce}) an odd and even number of times. 
The Appendix focuses on general correlation inequalities that hold for the monomer-dimer model on trees.

\section{Definitions and general properties of the monomer-dimer model}
Let $G=(V,E)$ be a finite simple graph with vertex set $V$ and edge set $E\subseteq\{uv\equiv\{u,v\}\,|\,u,v\in V,\,u\neq v\}\,$.

\begin{df}
A \textit{dimeric configuration} on the graph $G$ is a family of edges $D\subseteq E$ no two of which have a vertex in common.
Given a dimeric configuration $D$, the associated \textit{monomeric configuration} is the set of free vertices:
\[\MM_G(D):=\{v\in V\,|\;\forall u\!\in\!V\;uv\notin D\}\, .\]
We say that the edges in the dimeric configuration $D$ are occupied by a \textit{dimer}, while the vertices in the monomeric configuration $\MM_G(D)$ are occupied by a \textit{monomer}. Notice that $|\MM_G(D)|=|V|-2\,|D|\,$.
\end{df}

\begin{figure}[h]
\centering
\includegraphics[scale=0.21]{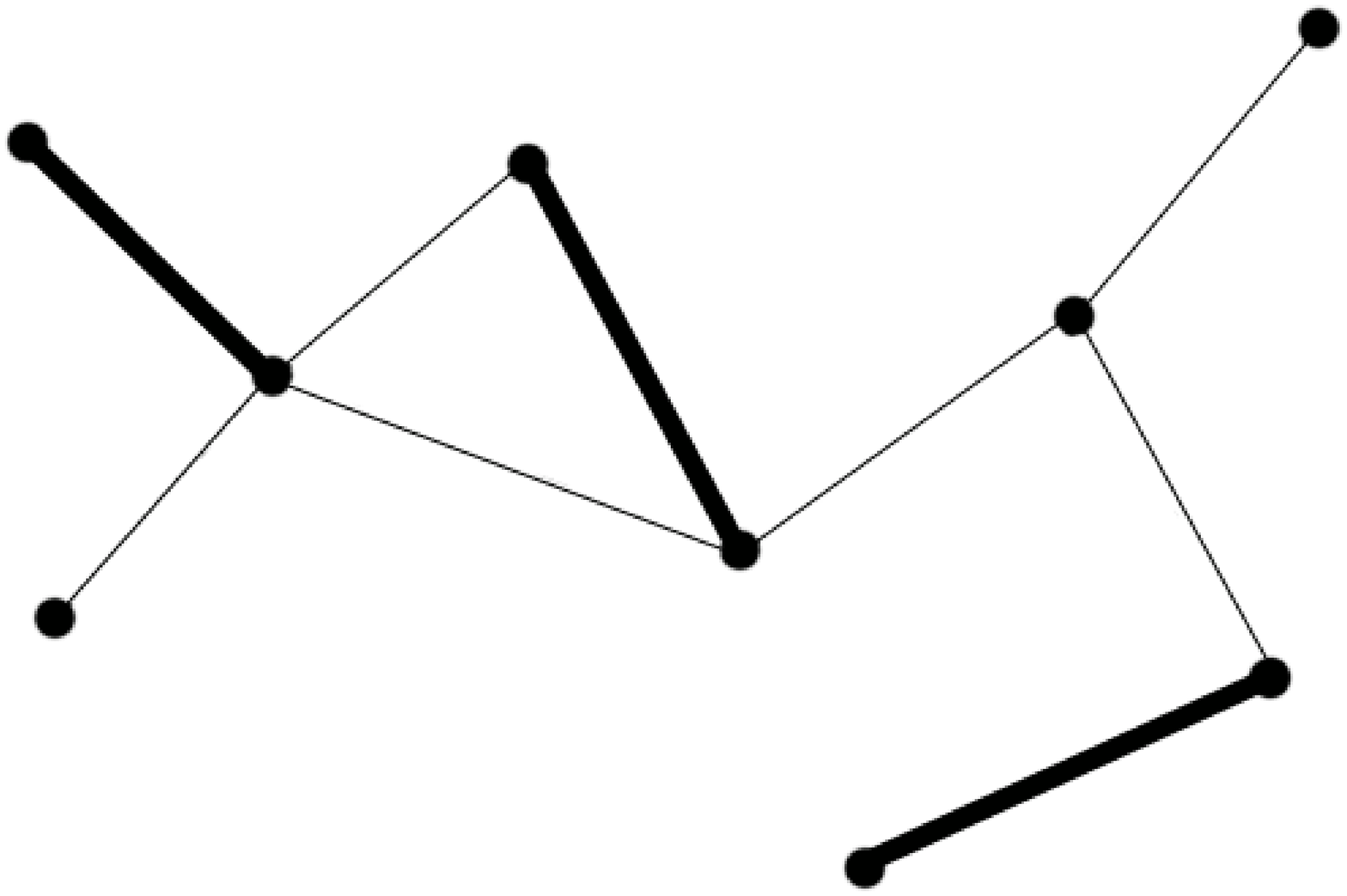} \qquad\quad
\includegraphics[scale=0.21]{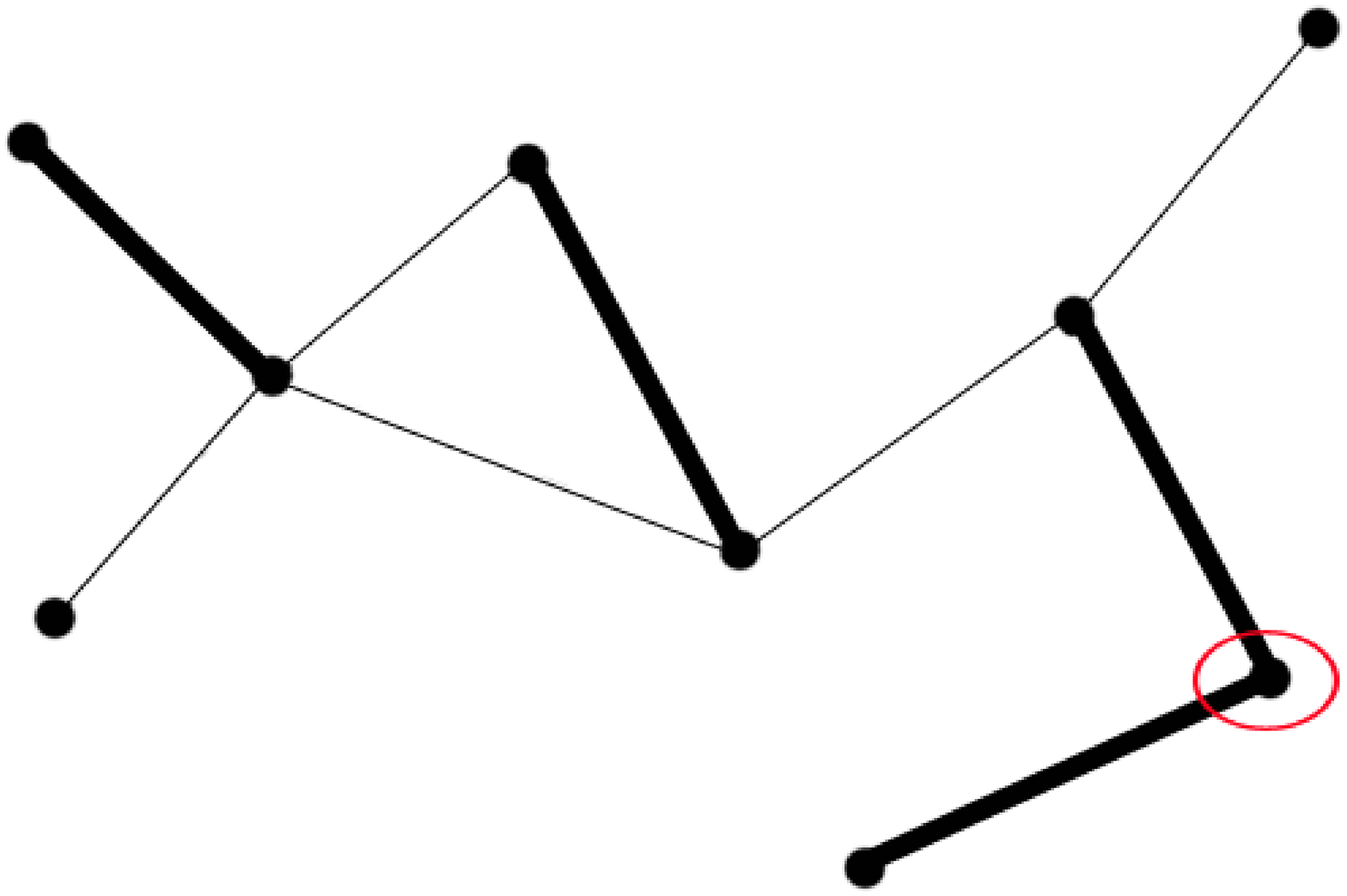}
\caption{The bold edges in the left figure form a dimeric configuration on the graph, while those in the right figure do not.}
\end{figure}

\begin{df}
Let $\DD_G$ be the set of all possible dimeric configurations on the graph $G\,$. 
For a given value of the parameter $x>0\,$, called \textit{monomer activity}, we define the following probability measure
on the set $\DD_G$:
$$\mu_{G,x}(D):=\frac{1}{Z_G(x)}\;x^{|V|-2\,|D|}\quad\forall D\!\in\!\DD_G\,.$$
The normalising factor
$$Z_G(x):=\sum_{D\in\DD_G}\,x^{|V|-2\,|D|}$$
is called \textit{partition function} of the model.
Its natural logarithm $\log Z_G$ is called \textit{pressure}.
The expected value with respect to the measure $\mu_{G,x}$ is denoted by $\bracket{\,\cdot\,}_{G,x}\,$, namely for any function $f$ of the dimeric configuration
$$\bracket{f}_{G,x}:=\sum_{D\in\DD_G}f(D)\,\mu_{G,x}(D)\, .$$
\end{df}

\begin{rk}\label{rk: general m-d model}
The general monomer-dimer model on the graph $G$ is obtained by assigning a monomeric weight $x_v>0$ to each vertex $v\in V$, a dimeric weight $w_e>0$ to each edge $e\in E$ and considering the measure
$$\mu_{G,\x,\w}(D)=\frac{1}{Z_G(\x,\w)}\;\tprod_{e\,\in D}w_e\,\tprod_{v\,\in\MM_G(D)}x_v \quad\forall D\!\in\!\DD_G\,.$$
In this paper we consider uniform monomeric and dimeric weights: $x_v\equiv x$, $w_e\equiv w$. Under this hypothesis one may assume without loss of generality $w=1$, indeed it's easy to check that
$$Z_G(x,w)\,=\,w^{|V|/2}\,Z_G\big(\frac{x}{\sqrt{w}}\,,1\big)\,.$$
\end{rk}

\begin{lem} \label{lem: pressure bounds}
The pressure per particle admits the following bounds:
$$ \log x \,\leq\, \frac{1}{|V|}\,\log Z_{G}(x) \,\leq\, \log x + \frac{|E|}{|V|}\,\log(1+\frac{1}{x^2})\;.$$
\end{lem}
\proof
The lower bound is obtained considering only the empty dimeric configuration (i.e. a monomer on each vertex of the graph):
$$ Z_G(x) \,\geq\, x^{|V|}\,.$$
The upper bound is obtained using the fact that any dimeric configuration made of $d$ dimers is a (particular) set of $d$ edges:
\[\begin{split}
Z_G(x) &\,=\, \sum_{d=0}^{|E|}\Card\{D\in\DD_G\, , \,|D|=d\}\;x^{|V|-2d} \,\leq\, \sum_{d=0}^{|E|}{|E|\choose d}\,x^{|V|-2d} \\
&\,=\, x^{|V|}\,(1+x^{-2})^{|E|}\;.\qedhere
\end{split}\]
\endproof

\begin{rk}\label{rk: monomer density}
An important quantity of the model is the expectation of the fraction of vertices covered by monomers.
A simple computation shows that it can be obtained from the pressure as:
\begin{equation}
\varepsilon_G(x) \,:=\, x\,\frac{\partial}{\partial x}\frac{\log Z_G(x)}{|V|} \,=\, \big\langle\frac{|\MM_G|}{|V|}\,\rangle_{G,x} \;.
\end{equation}
We call this quantity \textit{monomer density}.
It is useful to introduce the following notation for the probability of having a monomer on a given vertex $o\in V$:
\[ \RR_x(G,o):=\bracket{\,\1_{o\in\MM_G}}_{G,x}\in[0,1]\,. \]
Now the monomer density can be rewritten as
\begin{equation}\label{eq: monomer density}
\varepsilon_G(x)\,=\, \frac{1}{|V|}\,\sum_{o\in V}\RR_x(G,o) \,.
\end{equation}
\end{rk}

Two vertices $u,v\in V$ are \textit{neighbours} in the graph $G$ if there is an edge $uv\in E$ connecting them: we write $u\sim v\,$.
Denoting by $E_o$ the set of edges which connect the vertex $o\in V$ to one of its neighbours, we define the graph
$G-o:=(V\smallsetminus o, E\smallsetminus E_o)$.\\
Following \cite{BLS} we introduce a recursion relation for the probability $\RR_x(\cdot)$ that will be extensively used in the sequel; this is a rewriting of the recursion relation for the partition function $Z_\cdot(x)$ present in \cite{HL}.

\begin{lem} \label{lemma: recurrence relations} 
The family of functions $\RR_x(G,o)$ fulfils the relation
\begin{equation} \label{eq: recursion1}
\RR_x(G,o) = \frac{x^2}{x^2+\sum_{v\sim o}\RR_x(G-o,v)}\\[2pt]
\end{equation}
\end{lem}

\proof
The dimeric configurations on $G$ having a monomer on the vertex $o$ coincide with the dimeric configurations on $G-o$. Instead the dimeric configurations on $G$ having a dimer on the edge $ov$ are in one-to-one correspondence with the dimeric configurations on $G-o-v$.
Therefore
\[\RR_x(G,o) \,=\, \frac{1}{Z_G(x)}\!\!\sum_{D\in\DD_G\atop\text{s.t.}\,o\in\MM_G(D)} \!\!\!\!x^{|\MM_G(D)|} \,=\, \frac{x\,Z_{G-o}(x)}{Z_G(x)}\;,\]
\[\begin{split}
Z_G(x) &\,= \!\!\!\!\sum_{D\in\DD_G\atop\text{s.t.}\,o\in\MM_G(D)} \!\!\!\!x^{|\MM_G(D)|} \,+\, \sum_{v\sim o} \ \sum_{D\in\DD_G\atop\text{s.t.}\,ov\in D} x^{|\MM_G(D)|} \,=\\
&\,=\,x\,Z_{G-o}(x) + \sum_{v\sim o} Z_{G-o-v}(x)\,.
\end{split}\]
Hence one finds:
\[\begin{split}
\RR_x(G,o) &\,=\, \frac{x\,Z_{G-o}(x)}{x\,Z_{G-o}(x) + \sum_{v\sim o} Z_{G-o-v}(x)} \,=\,
\big(1+\tsum_{v\sim o} \dfrac{Z_{G-o-v}(x)}{x\,Z_{G-o}(x)}\,\big)^{-1}\\[2pt]
&\,=\, \big(1+x^{-2}\tsum_{v\sim o}\RR_x(G-o,v)\big)^{-1} \,=\,
\dfrac{x^2}{x^2+\sum_{v\sim o}\RR_x(G-o,v)}\,.\qedhere
\end{split}\]
\endproof

Iterating the recursion relation (\ref{eq: recursion1}), one obtains immediately the {\it squared} recursion relation 
\begin{equation} \label{eq: recursion2}
\RR_x(G,o) = \big(1+\tsum_{v\sim o}\dfrac{1}{x^2+\sum_{u\sim v,\,u\neq o}\RR_x(G-o-v,u)}\,\big)^{-1}\\[2pt] \, .
\end{equation}

In the next lemma we allow the monomer activity to take complex values, precisely those of the open half-plane
\[\HH_+ = \{z\in\C\,|\,\Re(z)>0\}\,.\]
This has no physical or probabilistic meaning, but it is a technique to obtain powerful results at real positive monomer activities by exploiting complex analysis.
This lemma already appeared in \cite{BLS} and in particular point $ii$ can be seen also as a consequence of theorem 4.2 in \cite{HL}.

\begin{lem}\label{lemma: complex weight}
\begin{itemize}
\item[i.] If $z\in\HH_+$, then $z^{-1}\,\RR_z(G,o)\in \HH_+$
\item[ii.] The function $z\mapsto\RR_z(G,o)$ is analytic on $\HH_+$
\item[iii.] If $z\in\HH_+$, then $|\RR_z(G,o)| \leq |z|/\Re(z)$
\end{itemize}
\end{lem}

\proof Note that $\HH_+$ is closed with respect to the operations $w\mapsto w^{-1}$ and $(w_1,w_2)\mapsto w_1+w_2$.\\
$[\mathbf{i,\,ii}]$ Proceed by induction on the number $N=|V|$ of vertices of the graph $G$.
For $N=1$ the graph $G$ coincides with its vertex $o$, hence $\RR_z(G,o)=1$. Therefore for $z\in\HH_+$, $z^{-1}\,\RR_z(G,o)=z^{-1}\in\HH_+$ and $\RR_z(G,o)\equiv1$ is obviously an analytic function of $z$.\\
Suppose now the statements \textit{i} and \textit{ii} hold for any graph of $N-1$ vertices and prove them for the graph $G$ of $N$ vertices.
By lemma \ref{lemma: recurrence relations}:
\[\RR_z(G,o) = \frac{z^2}{z^2+\sum_{v\sim o}\RR_z(G-o,v)} \,= \frac{z}{z+\sum_{v\sim o}z^{-1}\,\RR_z(G-o,v)}\]
By inductive hypothesis, for $z\in\HH_+$ and for every $v\sim o$, $z^{-1}\,\RR_z(G-o,v)\in\HH_+$ and $\RR_z(G-o,v)$ is an analytic  function of $z$. Therefore
\[z+\tsum_{v\sim o}z^{-1}\,\RR_z(G-o,v) \in\HH_+ \quad\text{(in particular it is $\neq0$)}\]
so that
$z^{-1}\RR_z(G,o)\in\HH_+$ and $\RR_z(G,o)$ is an analytic function of $z$ (as it is the quotient of non-zero analytic functions).\\
$[\mathbf{iii}]$ Use lemma \ref{lemma: recurrence relations}, then apply the elementary inequality $|z+w|\geq\Re(z+w)$ and conclude using point\textit{ i}:
\[\begin{split}
|\RR_z(G,o)| &= \big|\frac{z}{z+\sum_{v\sim o}z^{-1}\,\RR_z(G-o,v)}\,\big| \leq \frac{|z|}{\Re(z)+\sum_{v\sim o}\underbrace{\Re\big(z^{-1}\,\RR_z(G-o,v)\big)}_{>\,0}} \\[-9pt]
&\leq \frac{|z|}{\Re(z)}\,.\\[-25pt]
\end{split}\]
\endproof

In the graph $G$, given $o\in V$ and $r\in\N$, we denote by $[G,o]_r$ the \textit{ball of center $o$ and radius $r$}, that is the (connected) subgraph of $G$ induced by the vertices at graph-distance $\leq r$ from the origin $o$.\\
A \textit{tree} is a connected graph with no cycles.
If the graph $G$ is locally a tree near the vertex $o$, the next lemma allows to bound the operator $\RR_x(\cdot,o)$ from above/below by cutting away the ``non-tree'' part of $G$ at even/odd distance from $o$.

\begin{lem}[Correlation inequalities on a locally tree-like graph] \label{lemma: localisation} $\ $\\[6pt]
If $[G,o]_{2r}$ is a tree, then $\RR_x(G,o)\leq\RR_x([G,o]_{2r},o)\,$.\\[6pt]
If $[G,o]_{2r+1}$ is a tree, then $\RR_x(G,o)\geq\RR_x([G,o]_{2r+1},o)\,$.
\end{lem}

\proof
Proceed by induction on the distance $r\in\N$ from the origin $o$.\\[2pt]
For $r=0$, the graph $[G,o]_0$ is the isolate vertex $o$ hence
\[\RR_x(G,o)\leq 1 = \RR_x([G,o]_0,o)\,.\]
Assume now the result holds for $2r$ and prove it for $2r+1$ (with $r\geq0$).\\[1pt]
Suppose $[G,o]_{2r+1}$ is a tree. Note that $[G,o]_{2r+1}-o = \bigsqcup_{v\sim o}[G-o,v]_{2r}\,$, where $[G-o,v]_{2r}$ is a tree for every $v\sim o$.\\
As in general $\RR_x(H,v)$ depends only on the connected component of the graph $H$ which contains the vertex $v$, it follows:\[\RR_x([G,o]_{2r+1}\!-o,v)=\RR_x([G-o,v]_{2r},v)\,.\]
And by the induction hypothesis
\[\RR_x(G-o,v)\leq\RR_x([G-o,v]_{2r},v)\,.\]
Then using lemma \ref{lemma: recurrence relations} two times one obtains:
\[\begin{split}
\RR_x(G,o) &= \frac{x^2}{x^2+\sum_{i\sim o}\RR_x(G-o,i)} \,\geq \frac{x^2}{x^2+\sum_{i\sim o}\RR_x([G-o,i]_{2r},i)} \\ &= \frac{x^2}{x^2+\sum_{i\sim o}\RR_x([G,o]_{2r+1}\!-o,i)} \,= \RR_x([G,o]_{2r+1},o)\,.
\end{split}\]
Induction from $2r-1$ to $2r$ (with $r\geq1$) is done analogously.
\endproof

\section{The model on a Galton-Watson tree}
\begin{df}
As already said, a \textit{tree} $T$ is a connected graph with no cycles.\\
A \textit{rooted tree} is a tree $T$ together with the choice of a vertex $\vn$, the \textit{root}.\\
This choice induces an order relation on the vertex set of $T$: the vertices which are neighbours of the root $\vn$ form the \textit{$1^{st}$ generation}, the vertices different from $\vn$ and neighbours of a vertex in the $1^{st}$ generation compose the \textit{$2^{nd}$ generation}, and so on.
Given a vertex $v$, its \textit{sons} (or its \textit{offspring}) are the vertices in the following generation which are neighbours of $v$.\\
For $r\in\N$ we denote $T(r)$ the sub-tree of $T$ induced by the vertices in the first $r$ generations, namely $T(r)=[T,\vn]_r$.
The tree $T$ is \textit{locally finite} if the $T(r)$'s are finite graphs for every $r\in\N$.
\end{df}

The next proposition describes the behaviour of our model on any finite tree. While in this context it may be shown 
to be an easy consequence of lemma \ref{lemma: localisation}, it is also a special case of a general set of correlation 
inequalities that hold on trees which we include in the Appendix.

\begin{prop}\label{prop: alternate monotonicity on tree}
Let $T$ be a locally finite tree rooted at $\vn$.
Consider the monomer-dimer model on the finite sub-trees $T(r),\,r\in\N$.
Then:
\begin{itemize}
\item[i.] $r\mapsto\RR_x(T(2r),\vn)$ is monotonically decreasing
\item[ii.] $r\mapsto\RR_x(T(2r+1),\vn)$ is monotonically increasing
\item[iii.] $\RR_x(T(2r),\vn)\geq\RR_x(T(2s+1),\vn)\quad\forall\,r,s\in\N$
\end{itemize}
\end{prop}

\proof Let $r,s\in\N$.\\
$\mathbf{[i]}$ Consider the graph $T(2r+2)$. Cutting at distance $2r$ from $o$, one obtains $[T(2r+2),\vn]_{2r}=T(2r)$ which is a tree. Hence by lemma \ref{lemma: localisation}
\[\RR_x(T(2r+2),\vn)\leq\RR_x(T(2r),\vn)\,.\]
$\mathbf{[ii]}$  Consider the graph $T(2r+3)$. Cutting at distance $2r+1$ from $o$, one obtains $[T(2r+3),\vn]_{2r+1}=T(2r+1)$ which is a tree. Hence by lemma \ref{lemma: localisation}
\[\RR_x(T(2r+3),\vn)\geq\RR_x(T(2r+1),\vn)\,.\]
$\mathbf{[iii]}$ Consider the graph $T(2r+1)$. Cutting at distance $2r$ from $o$, one obtains $[T(2r+1),\vn]_{2r}=T(2r)$ which is a tree. Hence by lemma \ref{lemma: localisation}
\[\RR_x(T(2r+1),\vn)\leq\RR_x(T(2r),\vn)\,.\]
Now if $r\leq s$, combining point \textit{i.} and this third inequality, one finds
\[\RR_x(T(2r),\vn)\geq\RR_x(T(2s),\vn)\geq\RR_x(T(2s+1),\vn)\,;\]
while if $s\leq r$, combining point \textit{ii.} and the third inequality, one finds
\[\RR_x(T(2s+1),\vn)\leq\RR_x(T(2r+1),\vn)\leq\RR_x(T(2r),\vn)\,.\qedhere\]
\endproof

As a consequence of proposition \ref{prop: alternate monotonicity on tree} we obtain that on any locally finite rooted tree there exist $\lim_{r\to\infty}\RR_x(T(2r),\vn)\,$, $\lim_{r\to\infty}\RR_x(T(2r+1),\vn)$ and moreover
\begin{multline*}
0 \,\leq\, \lim_{r\to\infty}\RR_x(T(2r+1),\vn) \,=\, \sup_{r\in\N}\RR_x(T(2r+1),\vn) \,\leq\\
\leq\, \inf_{r\in\N}\RR_x(T(2r),\vn) \,=\, \lim_{r\to\infty}\RR_x(T(2r),\vn) \,\leq\, 1\,.
\end{multline*}

A natural question is if these two limits coincide or not.
In the next proposition we prove that they are analytic functions of the monomer activity $x$, so that it suffices to show that they coincide on a set of $x$'s admitting a limit point to conclude that they coincide for all $x>0$.
We first state the following lemma of general usefulness.

\begin{lem}\label{lemma: general property of bounded analytic}
Let $(f_n)_{n\in\N}$ be a sequence of complex analytic functions on $U\subseteq\C$ open. Suppose that\\
$\bullet$ for every compact $K\subset U$ there exists a constant $C_K<\infty$ such that
\[\sup_{z\in K}|f_n(z)| \leq C_K \quad \forall\,n\in\N\,;\]
$\bullet$ there exist $U_0\subseteq U$ admitting a limit point and a function $f$ on $U_0$ such that
\[f_n(z)\xrightarrow[n\to\infty]{}f(z) \quad \forall\,z\in U_0\,.\]
Then $f$ can be extended on $U$ in such a way that
\[f_n(z)\xrightarrow[n\to\infty]{}f(z) \quad \forall\,z\in U\,;\]
further the convergence is uniform on compact sets and $f$ is analytic on $U$.
\end{lem}

\proof
By hypothesis $(f_n)_{n\in\N}$ is a family of complex analytic functions on $U$, which is uniformly bounded on every compact subset $K\subset U$.
Therefore by Montel's theorem (e.g. see theorems 2.1 p. 308 and 1.1 p. 156 in \cite{L}), each sub-sequence $(f_{n_m})_{m\in\N}$ admits a further sub-sequence $(f_{n_{m_p}})_{p\in\N}$ that uniformly converges on every compact subset $K\subset U$ to an analytic function $f^{(\sigma)}$, where $\sigma=(n_{m_p})_{p\in\N}\,$.\\
On the other hand by the second hypothesis one already knows that
\[\forall\,z\in U_0\ \ \exists\lim_{n\to\infty}f_n(z)\,.\\[-3pt]\]
Thus by uniqueness of the limit, all the $f^{(\sigma)}$'s coincide on $U_0\,$.
Hence, as $U_0$ admits a limit point in $U$, by uniqueness of analytic continuation all the $f^{(\sigma)}$'s coincide on the whole $U$.
Denoting $f$ their common value, this entails that
\[\forall\,z\!\in\!U\ \ \exists\lim_{n\to\infty}f_n(z)=f(z)\,.\qedhere\]
\endproof

\begin{prop}\label{prop: analyticity}
Let $T$ be a locally finite tree rooted at $\vn$. Consider the monomer-dimer model on the sub-trees $T(r),\,r\in\N$.
Then
the maps
\[x\mapsto \lim_{r\to\infty}\RR_x(T(2r),\vn) \quad , \quad x\mapsto \lim_{r\to\infty}\RR_x(T(2r+\!1),\vn)\]
are analytic on $\R_+$.
\end{prop}

\proof
Set $f_r(z):=\RR_z(T(2r),o)$ and $g_r(z):=\RR_z(T(2r+1),o)$.
By lemma \ref{lemma: complex weight} $(f_r)_{r\in\N}$ is a family of complex analytic functions on $\HH_+$, and it is uniformly bounded on every compact subset $K\subset\HH_+$:
\[\begin{split}
\sup_{z\in K}|f_r(z)| \,\leq\, \sup_{z\in K}\frac{|z|}{\Re(z)} \,< \infty \quad\forall\,r\in\N\,.\\[-4pt]
\end{split}\]
On the other hand by proposition \ref{prop: alternate monotonicity on tree} one already knows that 
\[\forall\,x\!>\!0\ \ \exists\lim_{r\to\infty}f_r(x)\,.\]
The result for $(f_r)_{r\in\N}$ then follows by lemma \ref{lemma: general property of bounded analytic}.
The same reasoning holds for the sequence $(g_r)_{r\in\N}$.
\endproof

Now we define an important class of random trees. We will prove that for these trees the previous limits on even and odd depth almost surely coincide at every monomer activity. 

\begin{df}
Let $P=(P_k)_{k\in\N}\,$, $\rho=(\rho_k)_{k\in\N}\,$ be two probability distributions over $\N\,$.
A \textit{Galton-Watson tree} $\T(P,\rho)$ is a random tree rooted at $\vn$ and defined constructively as follows.\\
Let $\Delta$ be a random variable with distribution $P$, let $(K_{r,i})_{r\geq1,\,i\geq1}$ be i.i.d. random variables with distribution $\rho$ and independent of $\Delta\,$.
\begin{itemize}
\item[1)] Connect the root $\vn$ to $\Delta$ offspring, which form the $1^{st}$ generation
\item[2)] Connect each node $(r,i)$ in the $r^{th}$ generation to $K_{r,i}$ offspring; all together these nodes form the $(r+1)^{th}$ generation
\end{itemize}
Repeat recursively the second step for all $r\geq1$ and obtain $\T(P,\rho)$.\\
We denote $\T(P,\rho,r)$ the sub-tree of $\T(P,\rho)$ induced by the first $r$ generations.
Note that $\T(P,\rho)$ is locally finite.\\[1pt]
A special case of Galton-Watson tree is when $\rho=P$, which we simply denote $\T(\rho):=\T(\rho,\rho)$ and $\T(\rho,r):=\T(\rho,\rho,r)\,$.\\[1pt]
If instead the offspring distributions satisfy $\Po:=\sum_{k=0}^\infty k\,P_k<\infty$ and
\[\rho_k=\frac{(k+1)\,P_{k+1}}{\Po} \quad\forall k\in\N\,,\]
we call $\T(P,\rho)$ a \textit{unimodular} Galton-Watson tree.
\end{df}

In the following when we consider a Galton-Watson tree we suppose it is defined on the probability space $(\Omega,\F,\Prob)$ and we denote $\E[\,\cdot\,]$ the expectation with respect to the measure $\Prob$.
It is important to notice that when the monomer-dimer model is studied on a random graph $G$, then the measure $\mu_{G,x}$ is a random measure and therefore the probability $\RR_x(G,o)$ is a random variable. 

\begin{thm}\label{th: Galton-Watson}
Let $\T(\rho)$ be a Galton-Watson tree such that $\rhoo:=\sum_{k\in\N}k\,\rho_k<\infty\,$.
Consider the monomer-dimer model on the finite sub-trees $\T(\rho,r),\,r\in\N$.
Then almost surely for every $x>0$
\[\exists\,\lim_{r\to\infty}\RR_x(\T(\rho,r),\vn) =:X(x)\,.\]
The random function $x\mapsto X(x)$ is almost surely analytic on $\R_+\,$.\\[2pt]
The distribution of the random variable $X(x)$ is the only solution supported in $[0,1]$ of the following fixed point distributional equation:
\begin{equation}\label{eq: fixed point 1}
X \overset{\cal D}{=} \frac{x^2}{x^2+\sum_{i=1}^K X_i}\;,
\end{equation}
where $(X_i)_{i\in\N}$ are i.i.d. copies of $X$, $K$ has distribution $\rho$, $(X_i)_{i\in\N}$ and $K$ are independent.
\end{thm}

\proof
To ease the notation we drop the symbol $\rho$ as $\T:=\T(\rho)$ and $\T(r):=\T(\rho,r)$. 
By proposition \ref{prop: alternate monotonicity on tree} there exist the two limits
$$X^+(x):=\lim_{r\to\infty}\RR_x(\T(2r),\vn)\ \ ,\ \ X^-(x):=\lim_{r\to\infty}\RR_x(\T(2r+1),\vn)\,,$$
moreover $0\leq X^-\leq X^+\leq 1$ and by proposition \ref{prop: analyticity} the functions $x\mapsto X^+(x)$ and $x\mapsto X^-(x)$ are analytic on $\R_+$.
The theorem is obtained by the following lemmata.
\begin{lem}\label{lem:theoremproof1}
Given $x>0$, $X^+(x)$ and $X^-(x)$ are both solutions of the following fixed point distributional equation:
\begin{equation}\label{eq: fixed point 2}
X \overset{\cal D}{=} \big(1+\tsum_{i=1}^K(x^2+\tsum_{j=1}^{H_i}X_{i,j})^{-1}\big)^{-1}\;,
\end{equation}
where $(X_{i,j})_{i,j\in\N}$ are i.i.d. copies of $X$, $(H_i)_{i\in\N}$ are i.i.d. with distribution $\rho$, $K$ has distribution $\rho$, $(X_{i,j})_{i,j\in\N}$, $(H_i)_{i\in\N}$ and $K$ are mutually independent.
\end{lem}
We will write $u\leftarrow v$ to denote ``$u$ son of $v$ in the rooted tree $(\T,\vn)$''.
We will indicate $\T_u(r)$ the sub-tree of $\T$ induced by the vertex $u$ and its descendants until the $r^{th}$ generation (starting counting from $u$).
Using lemma \ref{lemma: recurrence relations} and precisely equation (\ref{eq: recursion2}) one finds, with the notations
just introduced,
\[\begin{split}
\RR_x(\T(2r+2),\vn) &= \big(1+\tsum_{v\leftarrow\vn}\big(x^2+\tsum_{u\leftarrow v}\RR_x(\T(2r+2)-\vn-v,u)\big)^{-1}\,\big)^{-1} \\[2pt]
&= \big(1+\tsum_{v\leftarrow\vn}\big(x^2+\tsum_{u\leftarrow v}\RR_x(\T_u(2r),u)\big)^{-1}\,\big)^{-1} \\[2pt]
&\overset{\cal D}{=} \big(1+\tsum_{i=1}^K\big(x^2+\tsum_{j=1}^{H_i}\RR_x(\T_{i,j}(2r),\vn)\big)^{-1}\,\big)^{-1}\,,
\end{split}\]
where $(\T_{i,j}(2r))_{i,j\in\N}$ are i.i.d. copies of $\T(2r)$, independent of $(H_i)_{i\in\N}$ and $K$.\\
Now since $\RR_x(\T(2r),\vn)\xrightarrow[r\to\infty]{a.s.}X^+(x)$, it holds also
\[\RR_x(\T(2r),\vn)\xrightarrow[r\to\infty]{\cal D}X^+(x)\,,\]
and moreover, thanks to the mutual independence of $\big(\RR_x(\T_{i,j}(2r),\vn)\big)_{i,j\in\N}\,$, $(H_i)_{i\in\N}\,$, $K\,$,
by standard probability arguments\footnote{equivalence between convergence in distribution and convergence of the characteristic functions (e.g. see theorems 26.3 p. 349 and 29.4 p. 383 in \cite{B0}) can be used.}
\[\big(\,\big(\RR_x(\T_{i,j}(2r),\vn)\big)_{i,j\in\N}\,,\,(H_i)_{i\in\N}\,,\,K\,\big) \xrightarrow[r\to\infty]{\cal D} \big(\,(X^+_{i,j})_{i,j\in\N}\,,\,(H_i)_{i\in\N}\,,\,K\,\big)\,,\]
where $(X^+_{i,j})_{i,j\in\N}$ are i.i.d. copies of $X^+(x)$, independent of $(H_i)_{i\in\N}$ and $K$.\\
Then for any bounded continuous function $\phi:[0,1]\rightarrow\R$
\[\begin{split}
\E[\phi(X^+(x))] &= \lim_{r\to\infty}\E[\phi\big(\RR_x(\T(2r+2),\vn)\big)] \\
&= \lim_{r\to\infty}\E\bigg[\phi\bigg(\big(1+\tsum_{i=1}^K\big(x^2+\sum_{j=1}^{H_i}\RR_x(\T_{i,j}(2r),\vn)\big)^{-1}\,\big)^{-1}\bigg)\bigg] \\
&= \E\bigg[\phi\bigg(\big(1+\tsum_{i=1}^K(x^2+\sum_{j=1}^{H_i}X^+_{i,j})^{-1}\,\big)^{-1}\bigg)\bigg]\;.
\end{split}\]
Namely $X^+(x)$ is a solution of distributional equation (\ref{eq: fixed point 2}).\\
In an analogous way it can be proven that also $X^-(x)$ is a solution of distributional equation (\ref{eq: fixed point 2}).
\begin{lem}\label{lem:theoremproof2}
Almost surely for all $x>0\,$ $X^+(x)=X^-(x)\,.$
\end{lem}
By proposition \ref{prop: alternate monotonicity on tree} $X^+(x)\geq X^-(x)$. By lemma \ref{lem:theoremproof1} $X^+(x)$ and $X^-(x)$ are both solutions of equation (\ref{eq: fixed point 2}). Therefore, taking $\big((H_i)_{i\in\N}\,,K\big)$ independent of $\big((X^+_{i,j})_{i,j\in\N}\,,\,(X^-_{i,j})_{i,j\in\N}\big)$, one obtains:
\[\begin{split}
& \E[|X^+(x)-X^-(x)|] \,=\, |\E[X^+(x)]-\E[X^-(x)]| \,=\, \\[4pt]
&=\, \big|\,\E\big[\big(1+\tsum_{i=1}^K(x^2+\sum_{j=1}^{H_i}X^+_{i,j})^{-1}\;\big)^{-1}\big]\ +\\ &\qquad\qquad\qquad\qquad\qquad\qquad\qquad\quad- \E\big[\big(1+\tsum_{i=1}^K(x^2+\sum_{j=1}^{H_i}X^-_{i,j})^{-1}\;\big)^{-1}\big]\big| \\[5pt]
&=\, \big|\,\E\big[ \frac{\sum_{i=1}^K(x^2+\sum_{j=1}^{H_i}X^-_{i,j})^{-1}-\sum_{i=1}^K(x^2+\sum_{j=1}^{H_i}X^+_{i,j})^{-1}} {\big(1+\sum_{i=1}^K(x^2+\sum_{j=1}^{H_i}X^+_{i,j})^{-1}\big)\big(1+\sum_{i=1}^K(x^2+\sum_{j=1}^{H_i}X^-_{i,j})^{-1}\big)} \big]\big| \\[5pt]
&=\, \big|\,\E\big[ \big(\tsum_{i=1}^K\dfrac{\sum_{j=1}^{H_i}(X^+_{i,j}-X^-_{i,j})}{(x^2+\sum_{j=1}^{H_i}X^-_{i,j})\,(x^2+\sum_{j=1}^{H_i}X^+_{i,j})}\big)\,\cdot\\[2pt] 
&\qquad\ \ \cdot\big(1+\tsum_{i=1}^K(x^2+\tsum_{j=1}^{H_i}X^+_{i,j})^{-1}\big)^{-1}\, \big(1+\tsum_{i=1}^K(x^2+\tsum_{j=1}^{H_i}X^-_{i,j})^{-1}\big)^{-1} \big]\big| \\[3pt]
&\leq\, \frac{1}{x^4}\;\E\big[\tsum_{i=1}^K\tsum_{j=1}^{H_i}|X^+_{i,j}-X^-_{i,j}|\,\big] \,=\, \dfrac{\rhoo^{\,2}}{x^4}\ \E[|X^+(x)-X^-(x)|\,]\;,
\end{split}\]
where the last equality is true by independence.\\
If $x>\sqrt{\rhoo}\,$, the contraction coefficient is $\rhoo^2/x^4<1$. Therefore for all $x>\sqrt{\rhoo}$
$$\E[|X^+(x)-X^-(x)|]=0\,,\quad\text{i.e.}\ \ X^+(x)=X^-(x)\;a.s.$$
As $\Q$ is countable it follows that
$$\big(X^+(x)=X^-(x)\ \forall\,x\in\,]\sqrt{\rhoo},\infty[\;\cap\,\Q\,\big)\ a.s.$$
Now remind that by proposition \ref{prop: analyticity} $X^+(x),\,X^-(x)$ are analytic functions of $x>0$.
Hence, as $\Q$ is dense in $\R$, this entails that
$$\big(X^+(x)=X^-(x)\ \forall\,x>0\,\big)\ a.s.$$
by uniqueness of the analytic continuation.\\
As a consequence $\big(\,\exists \lim_{r\to\infty}\RR_x(\T(r),\vn)=X^+(x)=X^-(x)\;\forall\,x>0\,\big)\ a.s.$
We call $X(x)$ this random analytic function of $x$.
\begin{lem}\label{lem:theoremproof3}
Given $x>0$, the random variable $X(x)$, satisfying the distributional equation (\ref{eq: fixed point 2}),
satisfies also the distributional equation (\ref{eq: fixed point 1}).
\end{lem}
Using lemma \ref{lemma: recurrence relations} and precisely equation (\ref{eq: recursion1}), one finds\\[-4pt]
\[\begin{split}
\RR_x(\T(r+1),\vn) &= \frac{x^2}{x^2+\sum_{v\leftarrow\vn}\RR_x(\T(r+1)-\vn,v)}\,
= \frac{x^2}{x^2+\sum_{v\leftarrow\vn}\RR_x(\T_v(r),v)} \\[1pt]
&\overset{\cal D}{=} \frac{x^2}{x^2+\sum_{i=1}^K\RR_x(\T_i(r),\vn)}\;,\\[4pt]
\end{split}\]
where $(\T_i(r))_{i\in\N}$ are i.i.d. copies of $\T(r)$, independent of $K$.\\[4pt]
Now since $\RR_x(\T(r),\vn)\xrightarrow[r\to\infty]{a.s.}X(x)\,$ (by definition, which is possible thanks to lemma \ref{lem:theoremproof2}), it holds also\\[-4pt]
\[\RR_x(\T(r),\vn)\xrightarrow[r\to\infty]{\cal D}X(x)\,,\]
and moreover, thanks to the independence of $\big(\RR_x(\T_i(r),\vn)\big)_{i\in\N}\,$, $K\,$,
\[\big(\,\big(\RR_x(\T_i(r),\vn)\big)_{i\in\N}\,,\,K\,\big) \xrightarrow[r\to\infty]{\cal D} \big(\,(X_i)_{i\in\N}\,,\,K\,\big)\,,\]
where $(X_i)_{i\in\N}$ are i.i.d. copies of $X(x)$, independent of $K$.\\[2pt]
Then for any bounded continuous function $\phi:[0,1]\rightarrow\R$\\[-4pt]
\[\begin{split}
\E[\phi(X(x))] &\,=\, \lim_{r\to\infty}\E[\phi\big(\RR_x(\T(r+1),\vn)\big)] 
\,=\, \lim_{r\to\infty}\E\big[\phi\big(\frac{x^2}{x^2+\sum_{i=1}^K\RR_x(\T_i(r),\vn)}\big)\big] \\[-2pt]
&\,=\, \E\big[\phi\big(\frac{x^2}{x^2+\sum_{i=1}^K X_i}\big)\big]\;.
\end{split}\]
Namely $X(x)$ is a solution of distributional equation (\ref{eq: fixed point 1}).
\begin{lem}\label{lem:theoremproof4}
For a given $x>0$, the distributional equation (\ref{eq: fixed point 1}) has a unique solution supported in $[0,1]$.
\end{lem}
Let $Y$ be a random variable taking values in $[0,1]$ and such that
$$Y \overset{\cal D}{=} \frac{x^2}{x^2+\sum_{i=1}^K Y_i}\;,$$
where $(Y_i)_{i\in\N}$ are i.i.d. copies of $Y$, independent of $K$. Observe that:
$$\begin{matrix}
\dfrac{x^2}{x^2+\sum_{i=1}^K Y_i} & \leq & 1 \\
\text{\begin{sideways}$\overset{\cal D}{=}\ $\end{sideways}} & & \text{\begin{sideways}$=\ $\end{sideways}} & \\
Y & & \RR_x(\T(0),\vn)
\end{matrix}$$
Therefore there exist $(Y_i')_{i\in\N}$ i.i.d. copies of $Y$ and $(\RR_x(\T(0),\vn)_i)_{i\in\N}$ i.i.d. copies of $\RR_x(\T(0),\vn)$ such that
\[Y_i' \,\leq\, \RR_x(\T(0),\vn)_i \quad\forall\,i\in\N\,.\\[-3pt]\]
Let $K'\overset{\cal D}{\sim}\rho$ independent of $(Y_i')_{i\in\N}\,$, $(\RR_x(\T(0),\vn)_i)_{i\in\N}\,$.
Applying the function $\frac{x^2}{x^2+\sum_{i=1}^{K'}(\,\cdot\,)}$, which is monotonically decreasing in each argument, to each term of the previous inequality one finds
\[\begin{split}\begin{matrix}
\dfrac{x^2}{x^2+\sum_{i=1}^{K'}\RR_x(\T(0),\vn)_i} & \leq & \dfrac{x^2}{x^2+\sum_{i=1}^{K'} Y_i'}\\
\text{\begin{sideways}$\overset{\cal D}{=}\ $\end{sideways}} & & \text{\begin{sideways}$\overset{\cal D}{=}\ $\end{sideways}}\\
\RR_x(\T(1),\vn) & & Y
\end{matrix}\end{split}\]
Therefore there exist $(\RR_x(\T(1),\vn)_i)_{i\in\N}$ i.i.d. copies of $\RR_x(\T(1),\vn)\,$ and $(Y_i'')_{i\in\N}$ i.i.d. copies of $Y$ such that
\[\RR_x(\T(1),\vn)_i \,\leq\, Y_i'' \quad\forall\,i\in\N\,.\\[-3pt]\]
Let $K''\overset{\cal D}{\sim}\rho$ independent of $(\RR_x(\T(1),\vn)_i)_{i\in\N}\,$, $(Y_i'')_{i\in\N}\,$.
Applying the function $\frac{x^2}{x^2+\sum_{i=1}^{K''}(\,\cdot\,)}$, which is monotonically decreasing in each argument, to each term of the previous inequality one finds
\[\begin{split}\begin{matrix}
\dfrac{x^2}{x^2+\sum_{i=1}^{K''} Y_i''} & \leq & \dfrac{x^2}{x^2+\sum_{i=1}^{K''}\RR_x(\T(1),\vn)_i'} \\
\text{\begin{sideways}$\overset{\cal D}{=}\ $\end{sideways}} & & \text{\begin{sideways}$\overset{\cal D}{=}\ $\end{sideways}} & \\
Y & & \RR_x(\T(2),\vn)
\end{matrix}\end{split}\]
Proceeding with this reasoning one obtains that for any $r\in\N$ there exist $\RR_x(\T(r),\vn)^{\sim} \overset{\cal D}{=} \RR_x(\T(r),\vn)$, $Y^{(r)} \overset{\cal D}{=} Y$ such that
\[\begin{split}\begin{matrix}
\RR_x(\T(2r+1),\vn)^{\sim} & \leq & Y^{(2r+1)} & \text{and} & Y^{(2r)}& \leq & \RR_x(\T(2r),\vn)^{\sim}\\
\text{\begin{sideways}$\xleftarrow[\phantom{\leftarrow r}]{\cal D}$\end{sideways}} & & & \text{as }r\to\infty & & &  \text{\begin{sideways}$\xleftarrow[\phantom{\leftarrow r}]{\cal D}$\end{sideways}}\\
X^-(x) & & & & & & X^+(x) &
\end{matrix}\\[-4pt]\end{split}\]
Since by lemma \ref{lem:theoremproof2} $X^+(x)=X^-(x)=X(x)\ a.s.$, it follows\footnote{A squeeze theorem for convergence in distribution holds: if $X_n\leq Y_n,\; Y_n'\leq X_n'$, $\,Y_n\overset{\cal D}{=}Y_n'\overset{\cal D}{=}Y$ for all $n\in\N$ and $X_n\xrightarrow[n\to\infty]{\cal D}X$, $\,X_n'\xrightarrow[n\to\infty]{\cal D}X$ then $Y\overset{\cal D}{=}X$.\\
To prove it work with the CDFs: $F_{X_n'}(x)\leq F_{Y_n'}(x)=F_{Y_n}(x)\leq F_{X_n}(x)\;\forall x\in\R$,
$F_{X_n'}(x)\xrightarrow[n\to\infty]{}F_{X}(x)$ and $F_{X_n}(x)\xrightarrow[n\to\infty]{}F_{X}(x)$ for every $x$ continuity point of $F_{X}$.
Since $F_{Y_n'}=F_{Y_n}=F_{Y}$, by the classical squeeze theorem it follows that $F_Y(x)=F_X(x)$ for every $x$ continuity point of $F_X$. Now since $F_X$ and $F_Y$ are right-continuous and the continuity points of $F_X$ are dense in $\R$, one concludes that $F_Y=F_X$.}
that $Y\overset{\cal D}{=}X(x)\,$. 
\endproof

\begin{cor}\label{cor: Galton-Watson}
Let $\T(P,\rho)$ be a Galton-Watson tree such that $\rhoo:=\sum_{k\in\N}k\,\rho_k<\infty\,$.
Consider the monomer-dimer model on the sub-trees $\T(P,\rho,r),\,r\in\N$.\\
Then almost surely for every $x>0$
\[\exists\,\lim_{r\to\infty}\RR_x(\T(P,\rho,r),\vn)=:Y(x)\,.\]
The random function $x\mapsto Y(x)$ is a.s. analytic on $\R_+$.\\[2pt]
The distribution of the random variable $Y(x)$ is
\[Y(x)\overset{\cal D}{=}\frac{x^2}{x^2+\sum_{i=1}^{\Delta}X_i}\,,\]
where $\Delta$ has distribution $P$ and is independent of $(X_i)_{i\in\N}\,$, $(X_i)_{i\in\N}$ are i.i.d. copies of $X$,  
the distribution of $X$ is the only solution supported in $[0,1]$ of the following fixed point distributional equation:
\[X \overset{\cal D}{=} \frac{x^2}{x^2+\sum_{i=1}^K X_i}\;,\]
where $K$ has distribution $\rho$ and is independent of $(X_i)_{i\in\N}$.
\end{cor}

\proof
We drop the symbols $P,\,\rho$ as $\T^*:=\T(P,\rho)$ and $\T^*(r):=\T(P,\rho,r)$.\\
Observe that $\T^*-\vn = \bigsqcup_{v\leftarrow\vn}\T^*_v$ and the random trees $(\T^*_v)_{v\leftarrow\vn}$ are i.i.d. Galton-Watson trees of the type $\T(\rho)$.
Using lemma \ref{lemma: recurrence relations}
\[\begin{split}
\RR_x(\T^*(r+1),\vn) &= \frac{x^2}{x^2+\sum_{v\leftarrow\vn}\RR_x(\T^*(r+1)-\vn,v)}\,
= \frac{x^2}{x^2+\sum_{v\leftarrow\vn}\RR_x(\T^*_v(r),v)} \\[1pt]
\end{split}\]
By theorem \ref{th: Galton-Watson} for any $v$ son of $\vn$, $\lim_{r\to\infty}\RR_x(\T^*_v(r),\vn)$ almost surely exists, it is analytic, and its distribution satisfies equation (\ref{eq: fixed point 1}).
Therefore $\lim_{r\to\infty}\RR_x(\T^*(r),\vn)$ almost surely exists and is analytic, in fact
\[\begin{split}
\lim_{r\to\infty}\RR_x(\T^*(r),\vn) &=  \frac{x^2}{x^2+\sum_{v\leftarrow\vn}\lim_{r\to\infty}\RR_x(\T^*_v(r),v)} \\
&\overset{\cal D}{=} \frac{x^2}{x^2+\sum_{i=1}^\Delta X_i}\;,
\end{split}\]
where $(X_i)_{i\in\N}$ are i.i.d. copies of the solution supported in $[0,1]$ of equation (\ref{eq: fixed point 1}) , $\Delta$ has distribution $P$ and is independent of $(X_i)_{i\in\N}\,$.
\endproof

\begin{cor}\label{cor: Galton-Watson complex}
In the hypothesis of corollary \ref{cor: Galton-Watson}, almost surely for every $z\in\HH_+$
\[\exists\,\lim_{r\to\infty}\RR_z(\T(P,\rho,r),\vn)=:Y(z)\,.\]
The random function $z\mapsto Y(z)$ is almost surely analytic on $\HH_+$ and the convergence is uniform on compact subsets of $\HH_+$.
\end{cor}

\proof
Set $f_r(z):=\RR_z(\T(P,\rho,r),\vn)$. By lemma \ref{lemma: complex weight} $(f_r)_{r\in\N}$ is a sequence of complex analytic functions on $\HH_+$, uniformly bounded on compact subsets.\\
On the other hand by corollary \ref{cor: Galton-Watson} $(f_r)_{r\in\N}$ a.s. converges pointwise on $\R_+$.\\
Then the result follows from lemma \ref{lemma: general property of bounded analytic}.
\endproof

\section{The model on random graphs locally convergent to a Galton-Watson tree}
Let $G_n=(V_n,E_n),\;n\in\N$ be a sequence of random finite simple graphs,
defined on the probability space $(\Omega,\F,\Prob)$.\\ 
We introduce now the main class of graphs studied in this paper. The idea is to fix a radius $r$ and draw a vertex $v$ uniformly at random from the graph $G_n$: for $n$ large enough we want the ball $[G_n,v]_r$ to be a (truncated) Galton-Watson tree with arbitrary high probability.

\begin{df}
The random graphs sequence $(G_n)_{n\in\N}$ \textit{locally converges} to the unimodular Galton-Watson tree $\T(P,\rho)$ if for any $r\in\N$ and for any $(T,o)$ finite rooted tree with at most $r$ generations
$$\frac{1}{|V_n|}\,\sum_{v\in V_n}\1\big(([G_n,v]_{r},v)\cong(T,o)\big)\ \xrightarrow[n\to\infty]{a.s.}\ \Prob\big((\T(P,\rho,r),\vn)\cong (T,o)\big)$$
Here $\cong$ denotes the isomorphism relation between rooted graphs.
\end{df}

\begin{rk}\label{rk: local convergence}
The following statements are equivalent:
\begin{itemize}
\item[i.] $(G_n)_{n\in\N}$ locally converges to $\T(P,\rho)$
\item[ii.] \textit{a.s.} for all $r\in\N$ and $(T,o)$ finite rooted tree with at most $r$ generations
\[\frac{1}{|V_n|}\,\sum_{v\in V_n}\1\big(([G_n,v]_r,v)\cong(T,o)\big)\ \xrightarrow[n\to\infty]{}\ \Prob\big((\T(P,\rho,r),\vn)\cong(T,o)\big)\]
\item[iii.] \textit{a.s.} for all $r\in\N$ and $F$ bounded function, invariant under rooted graph isomorphisms,
\[\frac{1}{|V_n|}\,\sum_{v\in V_n} F\big([G_n,v]_r,v\big)\;\1([G_n,v]_r\,\text{is a tree})\ \xrightarrow[n\to\infty]{}\ \E\big[F\big(\T(P,\rho,r),\vn\big)\big]\]
\item[iv.] \textit{a.s.} for all $r\in\N$ and $(B,o)$ finite rooted graph with radius $\leq r$
\[\frac{1}{|V_n|}\,\sum_{v\in V_n}\1\big(([G_n,v]_r,v)\cong(B,o)\big)\ \xrightarrow[n\to\infty]{}\ \Prob\big((\T(P,\rho,r),\vn)\cong(B,o)\big)\]
\item[v.] \textit{a.s.} for all $r\in\N$ and $F$ bounded function, invariant under rooted graph isomorphisms,
\[\frac{1}{|V_n|}\,\sum_{v\in V_n} F\big([G_n,v]_r,v\big)\ \xrightarrow[n\to\infty]{}\ \E\big[F\big(\T(P,\rho,r),\vn\big)\big]\]
\end{itemize}
\end{rk}

\proof 
Let $\G(r)$, $\TT(r)$ be respectively the set of finite rooted \textit{graphs}, \textit{trees} with radius $\leq r$, considered up to isomorphism.
It is important to note that they are \textit{countable} sets.
In particular let $\G_d(r)$, $\TT_d(r)$ be respectively the set of finite rooted \textit{graphs}, \textit{trees} with radius $\leq r$ and maximum degree $\leq d$, and observe that
\begin{itemize}
\item $\G_d(r)$ and $\TT_d(r)$ are finite, indeed they contains only graphs with at most $1+d+(d-1)^2+\dots+(d-1)^r$ vertices
\item $\G_d(r)\subseteq\G_{d+1}(r)$ and $\TT_d(r)\subseteq\TT_{d+1}(r)\,$,
\item $\G(r)=\bigcup_{d\in\N}\G_d(r)\,$ and $\TT(r)=\bigcup_{d\in\N}\TT_d(r)\,$.
\end{itemize}
We are interested in the two following probability measures on $\G(r)$
\begin{gather*}
\nu_{r,n}(B,o)\,:=\,\frac{1}{|V_n|}\,\sum_{v\in V_n}\1\big(([G_n,v]_r,v)\cong(B,o)\big) \quad\forall\,(B,o)\in\G(r)\,,\\[2pt]
\nu_r(B,o)\,:=\,\Prob\big((\T(P,\rho,r),\vn)\cong(B,o)\big) \quad\forall\,(B,o)\in\G(r)\,.
\end{gather*}
Note that $\nu_{r,n}$ is a random measure since it is an empirical average over the balls of the random graph $G_n$. Fixed an elementary event $\omega\in\Omega$, we write $\nu_{r,n}^\omega$ for the corresponding deterministic measure.
Note instead that $\nu_r$ is a deterministic measure supported on $\TT(r)$.\\[2pt]
%
$[\mathbf{i\Rightarrow ii}]$
By hypothesis $i$ for all $r\in\N$ and $(T,o)\in\TT(r)$ there exists a measurable set $N_{r,(T,o)}$ such that $\Prob(N_{r,(T,o)})=0$ and
\[\nu_{r,n}^\omega(T,o)\xrightarrow[n\to\infty]{}\nu_r(T,o)\ \ \forall\,\omega\in\Omega\smallsetminus N_{r,(T,o)}\,.\]
As $\bigcup_{r\in\N}\TT(r)$ is countable, setting $N:=\bigcup_{r\in\N}\bigcup_{(T,o)\in\TT(r)}N_{r,(T,o)}$ we obtain that $\Prob(N)=0$ and
\[\nu_{r,n}^\omega(T,o)\xrightarrow[n\to\infty]{}\nu_r(T,o)\ \ \forall\,(T,o)\in\!\TT(r)\ \ \forall\,r\!\in\N\ \ \forall\,\omega\in\Omega\smallsetminus N\,.\]
$[\mathbf{ii\Rightarrow iii}]$
By hypothesis $ii$ there exists $N$ with $\Prob(N)=0$ such that for all $\omega\in\Omega\smallsetminus N$, $r\in\N$, $(T,o)\in\TT(r)$
\[\nu_{r,n}^\omega(T,o)\xrightarrow[n\to\infty]{}\nu_r(T,o)\,.\]
Now let $\omega\in\Omega\smallsetminus N$, $r\in\N$ and $F:\G(r)\rightarrow\R$ bounded.
Summing over $\T_d(r)$ which is finite, clearly:
\[\sum_{(T,o)\in\TT_d(r)}\!\!\!\!F(T,o)\,\nu_{r,n}^\omega(T,o)\ \xrightarrow[n\to\infty]{} \sum_{(T,o)\in\TT_d(r)}\!\!\!\!F(T,o)\,\nu_r(T,o)\,.\]
On the other hand the sum over the countable set $\TT'_d(r):=\TT(r)\smallsetminus\TT_d(r)$ is:
\[\begin{split}
&\big|\!\!\sum_{\,(T,o)\in\TT'_d(r)}\!\!\!\!F(T,o)\,\nu_{r,n}^\omega(T,o)\,\big| \,\leq\,
\sup|F|\!\!\!\sum_{\,(T,o)\in\TT'_d(r)}\!\!\!\!\nu_{r,n}^\omega(T,o) \,\leq\\[2pt]
&\sup|F|\;\big(1-\!\!\!\!\!\sum_{(T,o)\in\TT_d(r)}\!\!\!\!\nu_{r,n}^\omega(T,o)\big)\ \xrightarrow[n\to\infty]{}\ 
\sup|F|\;\big(\,1-\!\!\!\!\!\sum_{(T,o)\in\TT_d(r)}\!\!\!\!\nu_r(T,o)\,\big)\ \xrightarrow[d\to\infty]{}\ 0\,,
\end{split}\]
where the limit in $n$ is done by hypothesis $ii$ and finiteness of $\TT_d(r)$, while the limit in $d$ is done by monotone convergence.
Similarly one finds that:
\[\begin{split}
&\big|\!\!\!\sum_{\,(T,o)\in\TT_d'(r)}\!\!\!\!F(T,o)\,\nu_r(T,o)\,\big| \,\leq\,
\sup|F|\!\!\!\sum_{\,(T,o)\in\TT_d'(r)}\!\!\!\!\nu_r(T,o) \,\leq\\[2pt]
&\sup|F|\;\big(\,1-\!\!\!\!\!\sum_{(T,o)\in\TT_d(r)}\!\!\!\!\nu_r(T,o)\,\big)\ \xrightarrow[d\to\infty]{}\ 0\,.
\end{split}\]
These tree facts prove (using triangular inequality and $\limsup$) that
\[\sum_{(T,o)\in\TT(r)}\!\!\!\!F(T,o)\,\nu_{r,n}^\omega(T,o)\ \xrightarrow[n\to\infty]{} \sum_{(T,o)\in\TT(r)}\!\!\!\!F(T,o)\,\nu_r(T,o)\,.\]
$[\mathbf{iii\Rightarrow iv}]$
By hypothesis $iii$ there exists $N$ with $\Prob(N)=0$ such that for all $\omega\in\Omega\smallsetminus N$, $r\in\N$, $F:\G(r)\rightarrow\R$ bounded
\[\sum_{(T,o)\in\TT(r)}\!\!\!\!F(T,o)\,\nu_{r,n}^\omega(T,o)\ \xrightarrow[n\to\infty]{} \sum_{(T,o)\in\TT(r)}\!\!\!\!F(T,o)\,\nu_r(T,o)\,.\]
Let $\omega\in\Omega\smallsetminus N$, $r\in\N$, $(T,o)\in\TT(r)$. Taking $F(\cdot)=\1(\cdot\cong(T,o))\,$, clearly
\[\nu_{r,n}^\omega(T,o)\ \xrightarrow[n\to\infty]{}\ \nu_r(T,o)\,.\]
Let instead $(B,o)\in\G(r)\smallsetminus\TT(r)$. Clearly $\nu_r(B,o)=0$ and on the other hand, taking $F\equiv1$,
\[\nu_{r,n}^\omega(B,o) \,\leq\, 1-\!\!\!\!\!\sum_{(T,o)\in\TT(r)}\!\!\!\!\nu_{r,n}^\omega(T,o)\ \xrightarrow[n\to\infty]{}\ 1-\!\!\!\!\!\sum_{(T,o)\in\TT(r)}\!\!\!\!\nu_r(T,o) \,=\, 0\,.\]
$[\mathbf{iv\Rightarrow v}]$
This proof is very similar to $ii\Rightarrow iii$.
By hypothesis $iv$, there exists $N$ with $\Prob(N)=0$ such that for all $\omega\in\Omega\smallsetminus N$, $r\in\N$, $(B,o)\in\G(r)$
\[\nu_{r,n}^\omega(B,o)\ \xrightarrow[n\to\infty]{}\ \nu_r(B,o)\,.\]
Now let $\omega\in\Omega\smallsetminus N$, $r\in\N$ and $F:\G(r)\rightarrow\R$ bounded.
Summing over $\G_d(r)$ which is finite, clearly:
\[\sum_{(B,o)\in\G_d(r)}\!\!\!\!F(B,o)\,\nu_{r,n}^\omega(B,o)\ \xrightarrow[n\to\infty]{} \sum_{(B,o)\in\G_d(r)}\!\!\!\!F(B,o)\,\nu_r(B,o)\,.\]
On the other hand the sum over the countable set $\G'_d(r):=\G(r)\smallsetminus\G_d(r)$ is:
\[\begin{split}
&\big|\!\!\sum_{\,(B,o)\in\G'_d(r)}\!\!\!\!F(B,o)\,\nu_{r,n}^\omega(B,o)\,\big| \,\leq\,
\sup|F|\!\!\!\sum_{\,(B,o)\in\G'_d(r)}\!\!\!\!\nu_{r,n}^\omega(B,o) \,=\\[2pt]
&\sup|F|\;\big(1-\!\!\!\!\!\sum_{(B,o)\in\G_d(r)}\!\!\!\!\nu_{r,n}^\omega(B,o)\big)\ \xrightarrow[n\to\infty]{}\ 
\sup|F|\;\big(\,1-\!\!\!\!\!\sum_{(B,o)\in\G_d(r)}\!\!\!\!\nu_r(B,o)\,\big)\ \xrightarrow[d\to\infty]{}\ 0\,,
\end{split}\]
where the limit in $n$ is done by hypothesis $iv$ and finiteness of $\G_d(r)$, while the limit in $d$ is done by monotone convergence.
Similarly one finds that:
\[\begin{split}
&\big|\!\!\!\sum_{\,(B,o)\in\G_d'(r)}\!\!\!\!F(B,o)\,\nu_r(B,o)\,\big| \,\leq\,
\sup|F|\!\!\!\sum_{\,(B,o)\in\G_d'(r)}\!\!\!\!\nu_r(B,o) \,\leq\\[2pt]
&\sup|F|\;\big(\,1-\!\!\!\!\!\sum_{(B,o)\in\G_d(r)}\!\!\!\!\nu_r(B,o)\,\big)\ \xrightarrow[d\to\infty]{}\ 0\,.
\end{split}\]
These tree facts prove (using triangular inequality and $\limsup$) that
\[\sum_{(B,o)\in\G(r)}\!\!\!\!F(B,o)\,\nu_{r,n}^\omega(B,o)\ \xrightarrow[n\to\infty]{} \sum_{(B,o)\in\G(r)}\!\!\!\!F(B,o)\,\nu_r(B,o)\,.\]
$[\mathbf{v\Rightarrow i}]$
By hypothesis $v$ there exists $N$ with $\Prob(N)=0$ such that for all $\omega\in\Omega\smallsetminus N$, $r\in\N$, $F:\G(r)\rightarrow\R$ bounded
\[\sum_{(B,o)\in\G(r)}\!\!\!\!F(B,o)\,\nu_{r,n}^\omega(B,o)\ \xrightarrow[n\to\infty]{} \sum_{(B,o)\in\G(r)}\!\!\!\!F(B,o)\,\nu_r(B,o)\,.\]
Let $\omega\in\Omega\smallsetminus N$, $r\in\N$, $(T,o)\in\TT(r)$. Taking $F(\cdot)=\1(\cdot\cong(T,o))\,$, clearly
\[\nu_{r,n}^\omega(T,o)\ \xrightarrow[n\to\infty]{}\ \nu_r(T,o)\,.\qedhere\]
\endproof

Observe that local convergence of random graphs $(G_n)_{n\in\N}$ to the random tree $\T(P,\rho)$ is, in measure theory language, $a.s.-$weak convergence of random measures $(\nu_{r,n})_{n\in\N}$ to the measure $\nu_r$ for all $r\in\N$.
From this point of view remark \ref{rk: local convergence} gives different characterisations of the weak convergence of measures, valid in general for measures defined on a discrete countable set (in particular the equivalences $ii\Leftrightarrow iii$ and $iv\Leftrightarrow v$ can be seen as consequences of the Portmanteau theorem, e.g. see theorem 2.1 p.$\!$ 16 in \cite{B}).

\begin{rk}
In a graph $G$ the degree of a vertex $v$, denoted $\deg_G(v)$, is the number of neighbours of $v$.
If $(G_n)_{n\in\N}$ locally converges to $\T(P,\rho)$, then $P$ is the \textit{empirical degree distribution} of $G_n$ in the limit $n\to\infty$.\\
Indeed the degree is a local function ($\deg_{G}(v)=\deg_{[G,v]_1}(v)$) and clearly an indicator function is bounded, hence by remark \ref{rk: local convergence} a.s. for every $k\in\N$
$$\frac{1}{|V_n|}\sum_{v\in V_n}\1(\deg_{G_n}\!(v)=k) \,\xrightarrow[n\to\infty]{}\,
\Prob(\deg_{\T(P,\rho)}\!(\vn)=k) \,=\, P_k\,.$$
\end{rk}

\begin{df}
The random graphs sequence $(G_n)_{n\in\N}$ is \textit{uniformly sparse} if
$$\lim_{l\to\infty}\,\limsup_{n\to\infty}\,\frac{1}{|V_n|}\sum_{v\in V_n}\deg_{G_n}(v)\,\1(\deg_{G_n}(v)\geq l) \,=\, 0\ \ a.s.$$
\end{df}

\begin{rk}\label{rk: uniform sparsity}
If $(G_n)_{n\in\N}$ is uniformly sparse and locally convergent to $\T(P,\rho)$, then
$$\frac{|E_n|}{|V_n|} \,\xrightarrow[n\to\infty]{}\, \frac{1}{2}\;\Po\ \ a.s.$$
To prove it we write two times the number of edges as the sum of all vertices' degrees
\[2\,\frac{|E_n|}{|V_n|} \,= \frac{1}{|V_n|}\sum_{v\in V_n}\deg_{G_n}(v)\,.\\[-4pt]\]
Then we fix $l\in\N$ and we split the right-hand sum in two parts, concerning respectively smaller and grater than $l$ degrees .
To the first part we can apply the local convergence hypothesis (remark \ref{rk: local convergence}):
\[\begin{split}
\frac{1}{|V_n|}\sum_{v\in V_n}\deg_{G_n}(v)\,\1(\deg_{G_n}(v)\leq l) &\xrightarrow[n\to\infty]{a.s.} \E[\,\deg_{\T(P,\rho)}(\vn)\,\1(\deg_{\T(P,\rho)}(\vn)\leq l)] \\
&\xrightarrow[l\to\infty]\, \E[\deg_{\T(P,\rho)}(\vn)] \,=\, \Po\,.
\end{split}\]
To the second part we apply the uniform sparsity hypothesis:
$$\lim_{l\to\infty}\,\limsup_{n\to\infty}\,\frac{1}{|V_n|}\sum_{v\in V_n}\deg_{G_n}(v)\,\1(\deg_{G_n}(v)\geq l\!+\!1) \,=\, 0\ \ a.s.$$
\end{rk}

\begin{ex}
An \textit{Erd\H{o}s-R\'enyi random graph} $G_n$ is a graph with $n$ vertices, where each pair of vertices is linked by an edge independently with probability $c/n$.
The sequence $(G_n)_{n\in\N}$ is uniformly sparse and locally converges to the unimodular Galton-Watson tree $\T(P,\rho)$ with $P=\rho=\textrm{Poisson}(c)$.
For proof and further examples see \cite{DMbraz,DM}.
\end{ex}

The next theorem describes the asymptotic behaviour of the monomer density along a sequence of graphs which locally converges to a Galton-Watson tree. In \cite{S} a similar result is expressed in the language of cavity method.

\begin{thm}\label{main theorem}
Let $(G_n)_{n\in\N}$ be a sequence of finite random graphs, which:
\begin{itemize}
\item[i.] is locally convergent to the unimodular Galton-Watson tree $\T(P,\rho)$;
\item[ii.] has asymptotic degree distribution $P$ with finite second moment (equivalently $\rhoo<\infty$).
\end{itemize}
Consider the monomer-dimer model on the graphs $G_n,\,n\in\N$.
Then almost surely for all $x>0$ the monomer density
\[ \varepsilon_{G_n}\!(x) \,=\, x\,\frac{\partial}{\partial x}\frac{\log Z_{G_n}\!(x)}{|V_n|} \,=\,
\frac{1}{|V_n|}\sum_{v\in V_n}\RR_x(G_n,v) \ \xrightarrow[n\to\infty]{}\ \E[Y(x)]\;.\]
The function $x\mapsto \E[Y(x)]$ is analytic on $\R_+$.\\[2pt]
The random variable $Y(x)$ is defined in corollary \ref{cor: Galton-Watson}, that is its distribution is:
\[Y(x)\overset{\cal D}{=}\frac{x^2}{x^2+\sum_{i=1}^{\Delta}X_i}\;,\]
where $\Delta$ has distribution $P$ and is independent of $(X_i)_{i\in\N}\,$, $(X_i)_{i\in\N}$ are i.i.d. copies of $X$,  
the distribution of $X$ is the only solution supported in $[0,1]$ of the following fixed point distributional equation:
\[X \overset{\cal D}{=} \frac{x^2}{x^2+\sum_{i=1}^K X_i}\;,\]
where $K$ has distribution $\rho$ and is independent of $(X_i)_{i\in\N}$.
\end{thm}

\proof
Set $\T^*:=\T(P,\rho)$ and $\T^*(r):=\T(P,\rho,r)$.\\ 
Let $r\in\N$ and $v\in V_n$. If $[G_n,v]_{2r+1}$ is a tree, then lemma \ref{lemma: localisation} permits to localize the problem:
\[\RR_x(G_n,v)\;\1([G_n,v]_{2r+1}\text{\small{ is a tree}})
\left\{\begin{array}{l}
\!\leq\,\RR_x([G_n,v]_{2r},v)\;\1([G_n,v]_{2r+1}\text{\small{ is a tree}}) \\[6pt]
\!\geq\,\RR_x([G_n,v]_{2r+1},v)\;\1([G_n,v]_{2r+1}\text{\small{ is a tree}})
\end{array}\right.\]
Now work with the right-hand bounds and take the averages over a uniformly chosen vertex $v$. First let $n\to\infty$ using the hypothesis of local convergence (see remark \ref{rk: local convergence}) and then let $r\to\infty$ using the results on Galton-Watson trees (corollary \ref{cor: Galton-Watson}) and dominated convergence: almost surely for all $x>0$
\[\begin{split}
&\frac{1}{|V_n|}\sum_{v\in V_n}\RR_x([G_n,v]_{2r},v)\;\1([G_n,v]_{2r+1}\text{\small{ is a tree}})\ \xrightarrow[n\to\infty]{}\\[2pt] 
&\E\big[\RR_x(\T^*(2r),\vn)\big]\ \underset{r\to\infty}{\searrow}\
\E[Y(x)]
\end{split}\]
and similarly
\[\begin{split}
&\frac{1}{|V_n|}\sum_{v\in V_n}\RR_x([G_n,v]_{2r+1},v)\;\1([G_n,v]_{2r+1}\text{\small{ is a tree}})\ \xrightarrow[n\to\infty]{}\\[2pt] 
&\E\big[\RR_x(\T^*(2r+1),\vn)\big]\ \underset{r\to\infty}{\nearrow}\ 
\E[Y(x)]\,.
\end{split}\]
On the other hand observe that $a.s.$ for all $x>0$
\[\begin{split}
&\big|\frac{1}{|V_n|}\sum_{v\in V_n}\RR_x(G_n,v) - \frac{1}{|V_n|}\sum_{v\in V_n}\RR_x(G_n,v)\;\1([G_n,v]_{2r+1}\text{\small{ is a tree}})\big| 
\,\leq\\ &\frac{1}{|V_n|}\sum_{v\in V_n}\big(1-\1([G_n,v]_{2r+1}\text{\small{ is a tree}})\,\big) \,\xrightarrow[n\to\infty]{}\,
1-\Prob(\T^*(2r+1)\text{\small{ is a tree}}) = 0\,.
\end{split}\]
Therefore one finds that almost surely for all $x>0$
\[\limsup_{n\to\infty}\frac{1}{|V_n|}\sum_{v\in V_n}\RR_x(G_n,v) \,\leq\, \E[Y(x)]\;;\quad
\liminf_{n\to\infty}\frac{1}{|V_n|}\sum_{v\in V_n}\RR_x(G_n,v) \,\geq\, \E[Y(x)]\;.\]
Namely there exists
$$\lim_{n\to\infty}\frac{1}{|V_n|}\sum_{v\in V_n}\RR_x(G_n,v) \,=\, \E[Y(x)]\ \ \forall\,x>0\ \ a.s.\\[-5pt]$$
Remembering remark \ref{rk: monomer density} and in particular the identity (\ref{eq: monomer density}) the proof is concluded, except for the analyticity of $x\mapsto\E[Y(x)]$ which will follow from the next corollary.
\endproof

\begin{cor}\label{cor: main theorem complex}
In the hypothesis of theorem \ref{main theorem}, almost surely for all $z\in\HH_+$
\[ \varepsilon_{G_n}\!(z) \,=\, z\,\frac{\dd}{\dd z}\frac{\log Z_{G_n}\!(z)}{|V_n|} \,=\,
\frac{1}{|V_n|}\sum_{v\in V_n}\RR_z(G_n,v) \ \xrightarrow[n\to\infty]{}\ \E[Y(z)]\;,\]
where  the random variable $Y(z)$ is defined in corollary \ref{cor: Galton-Watson complex}.\\[2pt]
The function $z\mapsto \E[Y(z)]$ is analytic on $\HH_+$ and the convergence is uniform on compact subsets of $\HH_+\,$.\\
As a consequence almost surely for all $k\geq1$ and $z\in\HH_+$
\[\frac{\dd^k}{\dd z^k}\,\frac{\log Z_{G_n}\!(z)}{|V_n|} \ \xrightarrow[n\to\infty]{}\ \frac{\dd^k}{\dd z^k}\,\frac{\E[Y(z)]}{z}\;.\]
\end{cor}

\proof
By lemma \ref{lemma: complex weight} $(\varepsilon_{G_n})_{n\in\N}$ is a sequence of complex analytic functions on $\HH_+$, which is uniformly bounded on compact subsets $K\subset\HH_+$:
\[ \sup_{z\in K}|\varepsilon_{G_n}(z)| \,\leq\, \frac{1}{|V_n|}\sum_{v\in V_n}\sup_{z\in K}\,|\RR_z(G_n,o)|  \,\leq\, \sup_{z\in K}\frac{|z|}{\Re(z)} \,< \infty \ \ \forall\,n\in\N\,.\]
On the other hand by theorem \ref{main theorem} $(\varepsilon_{G_n}(x))_{n\in\N}$ a.s. converges pointwise on $\R_+$ to $\E[Y(x)]$.
Then lemma \ref{lemma: general property of bounded analytic} applies: $\E[Y(z)]$ is analytic in $z\in\HH_+$ and a.s.
\[ \varepsilon_{G_n}(z) \xrightarrow[n\to\infty]{} \E[Y(z)] \quad\text{uniformly in $z\in K$ for every compact $K\subset\HH_+$}\,\\[-4pt].\]
This entails also convergence of derivatives (e.g. see theorem 1.2 p.$\!$ 157 in \cite{L}).
\endproof

The existence and analyticity of the monomer density in the thermodynamic limit entails the same properties for the pressure per particle. Only the additional assumption of uniform sparsity is required.

\begin{cor}\label{cor:pressure limit}
Let $(G_n)_{n\in\N}$ be a sequence of random graphs, which:
\begin{itemize}
\item[i.] is locally convergent to the unimodular Galton-Watson tree $\T(P,\rho)$;
\item[ii.] has asymptotic degree distribution $P$ with finite second moment;
\item[iii.] is uniformly sparse.
\end{itemize}
Then almost surely for every $x>0$
$$\frac{1}{|V_n|}\,\log Z_{G_n}\!(x)\ \xrightarrow[n\to\infty]\,\ p(a)+\int_{a}^x\frac{\E[Y(t)]}{t}\;\dd t$$
where $a>0$ is arbitrary, $p(a)=\lim_{n\to\infty}\frac{1}{|V_n|}\,\log Z_{G_n}(a)$ a.s., and $Y(t)$ is the random variable defined in theorem \ref{main theorem}.\\[2pt]
The function $x\mapsto p(a)+\int_{a}^x\frac{\E[Y(t)]}{t}\;\dd t$ is analytic on $\R_+\,$.
\end{cor}

\proof
From theorem \ref{main theorem}, using the fundamental theorem of calculus and dominated convergence, it follows immediately that $a.s.$ for every $x>0,\,a>0$
\begin{equation}\label{eq:pressure difference limit}
\frac{\log Z_{G_n}(x)}{|V_n|}-\frac{\log Z_{G_n}(a)}{|V_n|} \,=\, \int_{a}^x \frac{\partial}{\partial t}\,\frac{\log Z_{G_n}(t)}{|V_n|}\,\dd t \,\xrightarrow[n\to\infty]{}\, \int_{a}^x\frac{\E[Y(t)]}{t}\,\dd t
\end{equation}
By theorem \ref{main theorem} the function $x\mapsto\E[Y(x)]$ is analytic on $\R_+$, therefore the integral function $x\mapsto\int_{a}^x\frac{\E[Y(t)]}{t}\,\dd t$ is analytic on $\R_+$ too.\\[2pt]
To conclude it remains to prove that almost surely for all $x>0$
$$\exists\,\lim_{n\to\infty}\frac{\log Z_{G_n}(x)}{|V_n|}\;.$$
Use the bounds for the pressure of lemma \ref{lem: pressure bounds} to estimate
\begin{equation}\label{eq:pressure difference bounds}
\frac{\log Z_{G_n}(x)}{|V_n|}-\frac{\log Z_{G_n}(a)}{|V_n|}\;
\left\{\begin{array}{l}
\leq\, \frac{\log Z_{G_n}(x)}{|V_n|}\,-\log a\\[8pt]
\geq\, \frac{\log Z_{G_n}(x)}{|V_n|}\,-\log a-\frac{|E_n|}{|V_n|}\,\log(1+\frac{1}{a^2})
\end{array}\right.
\end{equation}
Put together (\ref{eq:pressure difference limit}), (\ref{eq:pressure difference bounds}), remind $|E_n|/|V_n|\xrightarrow[n\to\infty]{a.s.}\Po/2$ and obtain that $a.s.$ for all $x>0$
\begin{gather*}
\liminf_{n\to\infty}\frac{\log Z_{G_n}(x)}{|V_n|} \;\geq\, \log a + \int_a^x \frac{\E[Y(t)]}{t}\,\dd t\;,\\[4pt]
\limsup_{n\to\infty}\frac{\log Z_{G_n}(x)}{|V_n|} \;\leq\, \log a + \frac{\Po}{2}\,\log(1+\frac{1}{a^2}) + \int_a^x \frac{\E[Y(t)]}{t}\,\dd t\;.
\end{gather*}
Therefore $a.s.$ for all $x>0$
$$0 \,\leq\, \limsup_{n\to\infty}\frac{\log Z_{G_n}(x)}{|V_n|} - \liminf_{n\to\infty}\frac{\log Z_{G_n}(x)}{|V_n|} \,\leq\, \frac{\Po}{2}\,\log(1+\frac{1}{a^2}) \,\xrightarrow[a\to\infty]\, 0\,,$$
which entails existence of $\lim_{n\to\infty}\frac{\log Z_{G_n}(x)}{|V_n|}$ and completes the proof.
\endproof

\begin{cor}
In the hypothesis of corollary \ref{cor:pressure limit}, if $\Po>0$,
almost surely the pressure
\[ \lim_{n\to\infty}\frac{\log Z_{G_n}}{|V_n|} \]
is an analytic function of the monomer density
\[ \lim_{n\to\infty}\varepsilon_{G_n}\;. \]
\end{cor}

\proof
Set $p_n:=\frac{\log Z_{G_n}}{|V_n|}$, $p:=\lim_{n\to\infty}p_n$ and $\varepsilon_n:=\varepsilon_{G_n}$, $\varepsilon:=\lim_{n\to\infty}\varepsilon_n$.\\
By theorem \ref{main theorem} and corollary \ref{cor:pressure limit} on an event of probability $1$ the monomer density $\varepsilon$ and the pressure $p$ are analytic functions of the monomer activity $x>0$.
Now a direct computation shows that
\[x\,\frac{\partial\varepsilon_n}{\partial x}(x) \,=\, \frac{<|\MM_{G_n}|^2>_{G_n,x}-<|\MM_{G_n}|>_{G_n,x}^2}{|V_n|} \,\geq\, 0\;.\]
But a more precise lower bound is provided by theorems 7.3 and 7.6 in \cite{HL}:
\[x\,\frac{\partial\varepsilon_n}{\partial x}(x) \,\geq\, \frac{|V_n|}{|E_n|}\;x^2\,\big(1-\varepsilon_n(x)\big)^2
\quad\textrm{and}\quad
1-\varepsilon_n(x) \,\geq\, \frac{2}{x^2+2}\;\frac{|E_n|}{|V_n|}\;,\]
hence
\[x\,\frac{\partial\varepsilon_n}{\partial x}(x) \,\geq\, \frac{4\,x^2}{(x^2+2)^2}\;\frac{|E_n|}{|V_n|}\ \xrightarrow[n\to\infty]{}\ \frac{2\,x^2}{(x^2+2)^2}\;\Po\;.\]
By corollary \ref{cor: main theorem complex} it follows:
\[x\,\frac{\partial\varepsilon}{\partial x}(x) \,\geq\, \frac{2\,x^2}{(x^2+2)^2}\;\Po \,>\,0\;.\]
Thus $\varepsilon$ is an analytic function of $x$ with non-zero derivative, so that it is invertible and its inverse is analytic (e.g. see theorem 6.1 p.$\!$ 76 of \cite{L}). 
In other words $x$ can be seen as an analytic function of $\varepsilon$.
Since the composition of analytic functions is analytic, it is proved that $p$ is an analytic function of $\varepsilon$.
\endproof

The following theorem improves corollary \ref{cor:pressure limit} giving an explicit expression of the asymptotic pressure. This has been found in \cite{AD}, using the heuristic of free energy shifts \cite{ZM} and then proving it is correct.

\begin{thm}\label{thm: pressure limit}
In the hypothesis of corollary \ref{cor:pressure limit}, almost surely for every $x>0$
\[ \frac{\log Z_{G_n}(x)}{|V_n|} \ \xrightarrow[n\to\infty]{}\ \E\big[\log\big(x+\sum_{i=1}^\Delta \frac{X_i(x)}{x}\,\big)\big] \,-\, \frac{\Po}{2}\;\E\big[\log\big(1+\frac{X_1(x)}{x}\;\frac{X_2(x)}{x}\big)\big] \]
where $\Delta$ has distribution $P$ and is independent of $(X_i)_{i\in\N}\,$, $(X_i)_{i\in\N}$ are i.i.d. copies of $X$,  
the distribution of $X$ is the only solution supported in $[0,1]$ of the fixed point distributional equation
\[X \overset{\cal D}{=} \frac{x^2}{x^2+\sum_{i=1}^K X_i}\;,\]
where $K$ has distribution $\rho$ and is independent of $(X_i)_{i\in\N}$.
\end{thm}

\proof
By theorem \ref{main theorem} and corollary \ref{cor:pressure limit} one already knows that almost surely there exist $\lim_{n\to\infty}x\,\frac{\partial}{\partial x}\frac{\log Z_{G_n}(x)}{|V_n|}=:\varepsilon(x)\,$ and $\lim_{n\to\infty}\frac{\log Z_{G_n}(x)}{|V_n|}=:p(x)\,$ and that
\begin{equation}\label{eq: pressure_theor1}
p(x) = p(a) + \int_a^x \frac{\varepsilon(t)}{t}\,\dd t\;,\quad \text{i.e. }\ x\;\frac{\partial p}{\partial x}(x) = \varepsilon(x)\;.
\end{equation}
Applying lemma \ref{lem: pressure bounds} to $G_n$ and passing to the limit exploiting remark \ref{rk: uniform sparsity}, one obtains the following bounds
\begin{equation}\label{eq: pressure_theor2}
\log x \,\leq\, p(x) \,\leq\, \log x + \frac{\Po}{2}\,\log(1+\frac{1}{x^2})\;,\quad \text{thus }\lim_{x\to+\infty}p(x)-\log x=0\;.
\end{equation}
Now set
\[ \widetilde{p}(x) \,:=\, \E\big[\log\big(x+\sum_{i=1}^\Delta \frac{X_i}{x}\big)\big] \,-\, \frac{\Po}{2}\;\E\big[\log\big(1+\frac{X_1}{x}\,\frac{X_2}{x}\big)\big]\; .\]
In order to prove that $p(x)=\widetilde{p}(x)$ it will suffice to show that $\widetilde{p}$ shares the two previous properties.
Hence split the proof in two lemmata.

\begin{lem}\label{lem: pressure_theor1}
For every $x>0$
\[ x\;\frac{\partial \widetilde{p}}{\partial x}\,(x) \,=\, \varepsilon(x) \;. \]
\end{lem}

The random complex function $z\mapsto X(z)=\lim_{r\to\infty}\RR_z(\T(\rho,r),o)$ is a.s. analytic on $\HH_+$ by corollary \ref{cor: Galton-Watson complex} and it is bounded by a deterministic function by lemma \ref{lemma: recurrence relations}: $|X(z)| \leq \frac{|z|}{\Re(z)}\,$.
As a consequence also its derivative at $z_0\in\HH_+$ is bounded by a deterministic constant, precisely fixing $r>0$ such that $\overline{B}(z_0,r)\subset\HH_+$ the integral representation (e.g. see theorem 7.3 p. 128 in \cite{L}) gives
\[\big|\frac{\dd X}{\dd z}(z_0)\big| \,=\, \big|\frac{1}{2\pi i}\int_{S(z_0,r)}\!\frac{X(z)}{(z-z_0)^2}\;\dd z\big| \,\leq\, \frac{1}{r}\,\max_{S(z_0,r)}\frac{|z|}{\Re (z)}\, =:c(z_0)\,. \]
It follows that the random functions under expectation in the expression of $\widetilde{p}$ are differentiable with integrable derivatives:
\[\big|x\,\frac{\partial}{\partial x}\log\big(x+\sum_{i=1}^\Delta\frac{X_i}{x}\big)\big| =
\big|\frac{x+\sum_{i=1}^\Delta (\frac{\partial X_i}{\partial x} - \frac{X_i}{x})}{x+\sum_{i=1}^\Delta\frac{X_i}{x}}\,\big| \leq
\frac{x+\Delta\,(c(x)+\frac{1}{x})}{x}\ \in L^1(\Prob),\]
\[\big|x\,\frac{\partial}{\partial x}\log\big(1+\frac{X_1}{x}\,\frac{X_2}{x}\big)\big| =
\big|\frac{\frac{\partial X_1}{\partial x}\,\frac{X_2}{x} + \frac{X_1}{x}\,\frac{\partial X_2}{\partial x} - 2\,\frac{X_1}{x}\,\frac{X_2}{x}}{1 + \frac{X_1}{x}\,\frac{X_2}{x}}\,\big| \leq
2c(x)\,\frac{1}{x} + 2\frac{1}{x^2}\,.\]
Thus one may apply Lebesgue's dominated convergence theorem and take the derivative under expectation, finding:
\[ x\;\frac{\partial \widetilde{p}}{\partial x}\,(x) \,=\,
\E\bigg[\frac{x+\sum_{i=1}^\Delta (\frac{\partial X_i}{\partial x} - \frac{X_i}{x})}{x+\sum_{i=1}^\Delta\frac{X_i}{x}}\bigg] \,-\, \frac{\Po}{2}\;\E\bigg[\frac{\frac{\partial X_1}{\partial x}\,\frac{X_2}{x} + \frac{X_1}{x}\,\frac{\partial X_2}{\partial x} - 2\,\frac{X_1}{x}\,\frac{X_2}{x}}{1 + \frac{X_1}{x}\,\frac{X_2}{x}}\bigg]\,. \]
Now reordering terms and setting
\begin{gather*}
I_0 \,:=\, \E\big[\frac{x}{x+\sum_{i=1}^\Delta\frac{X_i}{x}}\big] \\[2pt]
I_1 \,:=\, -\,\E\big[\frac{\sum_{i=1}^\Delta \frac{X_i}{x}}{x+\sum_{i=1}^\Delta \frac{X_i}{x}}\big] \,+\,
\Po\;\E\big[\frac{\frac{X_1}{x}\,\frac{X_2}{x}}{1+\frac{X_1}{x}\,\frac{X_2}{x}}\big] \\[2pt]
I_2 \,:=\, \E\big[\frac{\sum_{i=1}^\Delta \frac{\partial X_i}{\partial x}}{x+\sum_{i=1}^\Delta \frac{X_i}{x}}\big] \,-\,
\Po\;\E\big[\frac{\frac{X_1}{x}\,\frac{\partial X_2}{\partial x}}{1+\frac{X_1}{x}\,\frac{X_2}{x}}\big]
\end{gather*}
one may write $x\,\frac{\partial \widetilde{p}}{\partial x} = I_0+I_1+I_2\,$.
Observe that $I_0=\varepsilon(x)$ by theorem \ref{main theorem}. Then showing that $I_1=I_2=0$ will prove the lemma.\\
Start proving that $I_1=0$. First condition on the values of $\Delta$, use the fact that $(X_i)_{i\in\N}$ are i.i.d. and independent of $\Delta$ and $K$, and exploit the hypothesis of unimodularity (i.e. $d\,P_d=\Po\rho_{d-1}\ \forall\,d\geq1$):
\[\begin{split}
&\E\big[\frac{\sum_{i=1}^\Delta \frac{X_i}{x}}{x+\sum_{i=1}^\Delta \frac{X_i}{x}}\big] \,=\,
\sum_{d=0}^\infty\sum_{i=1}^d\,\E\big[\frac{\frac{X_i}{x}}{x+\sum_{i=1}^d \frac{X_i}{x}}\big]\,P_d \,=\,
\sum_{d=0}^\infty d\;\E\big[\frac{\frac{X_d}{x}}{x+\sum_{i=1}^d \frac{X_i}{x}}\big]\,P_d \\[2pt]
&=\,\sum_{d=1}^\infty \Po\;\E\big[\frac{\frac{X_d}{x}}{x+\sum_{i=1}^d \frac{X_i}{x}}\big]\,\rho_{d-1} \,=\,
\Po\;\E\big[\frac{\frac{X_{K+1}}{x}}{x+\sum_{i=1}^{K+1} \frac{X_i}{x}}\big]\,,
\end{split}\]
then exploit the fact that $X/x \overset{\cal D}{=} (x+\sum_{i=1}^K X_i/x)^{-1}$:
\[ \Po\;\E\big[\frac{\frac{X_{K+1}}{x}}{x+\sum_{i=1}^{K+1} \frac{X_i}{x}}\big] \,=\,
\Po\;\E\big[\frac{\frac{X_2}{x}}{(\frac{X_1}{x})^{-1} +\frac{X_2}{x}}\big] \,=\,
\Po\;\E\big[\frac{\frac{X_1}{x}\frac{X_2}{x}}{1+\frac{X_1}{x}\frac{X_2}{x}}\big]\,. \]
This proves $I_1=0$.
An analogous reasoning proves that $I_2=0$; one should only observe that the family of couples $(X_i\,,\,\frac{\partial X_i}{\partial x})_{i\in\N}$ can be chosen i.i.d. and independent of $\Delta$ and $K$ (it suffices to work on i.i.d. trees $(\T(\rho)_i)_{i\in\N}$).

\begin{lem}\label{lem: pressure_theor2}
\[ \lim_{x\to+\infty} \widetilde{p}(x)-\log x = 0 \;.\]
\end{lem}

A direct computation and the dominated convergence theorem give
\[ \widetilde{p}(x)-\log x =\,
\E\big[\log\big(1+\sum_{i=1}^\Delta \frac{X_i}{x^2}\big)\big] - \frac{\Po}{2}\;\E\big[\log\big(1+\frac{X_1\,X_2}{x^2}\big)\big] \xrightarrow[x\to\infty]{}\,
0 \]
indeed the function $x\mapsto X(x)$ is bounded in $[0,1]$ and for any $x\geq1$
\begin{gather*}
0 \leq \log\big(1+\sum_{i=1}^\Delta\,\frac{X_i}{x^2}\big) \leq \log(1+\Delta) \leq \Delta\ \in L^1(\Prob) \;, \\
0 \leq \log(1+\frac{X_1\,X_2}{x^2}) \leq \log 2 \;.
\end{gather*}

Now lemmata \ref{lem: pressure_theor1}, \ref{lem: pressure_theor2} together with formulae (\ref{eq: pressure_theor1}), (\ref{eq: pressure_theor2}) allow immediately to conclude the proof of the theorem:
\[\begin{split}
&p(x)-p(a) \,= \int_a^x \frac{\varepsilon(t)}{t}\,\dd t \,=\, \widetilde{p}(x)-\widetilde{p}(a)\ \ \Rightarrow\\[2pt]
&p(x)\underbrace{-p(a)+\log a}_{\xrightarrow[a\to\infty]{}\ 0} \,=\, \widetilde{p}(x)\underbrace{-\widetilde{p}(a)+\log a}_{\xrightarrow[a\to\infty]{}\ 0}\ \ \Rightarrow\ \ 
p(x)=\widetilde{p}(x)\;.\qedhere
\end{split}\]
\endproof

\section{Upper and lower bounds}
To conclude we consider the particular case when the graphs sequence $(G_n)_{n\in\N}$ locally converges to $\T(P,\rho)$ with $P=\rho=\textrm{Poisson}(2)$ (e.g. this is the case of $G_n$ Erd\H{o}s-R\'enyi with $c=2$), and we show an approximate plot of the monomer density $\varepsilon(x):=\displaystyle\lim_{n\to\infty}\varepsilon_{G_n}(x)$.

\begin{figure}[h]
\centering
\includegraphics[scale=0.44]{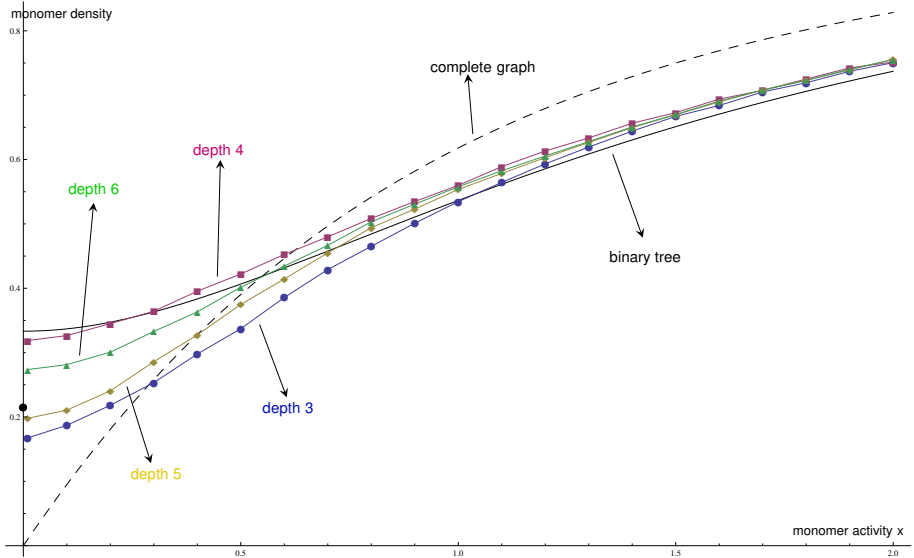}
\caption{Upper and lower bounds for the monomer density $\varepsilon$ on the Erd\H{o}s-R\'enyi with $c=2$ versus the monomer activity $x$ (squares, triangles, diamonds, circles). The monomer density on the binary tree (continuous line) and the complete graph (dashed line) versus the monomer activity, see \cite{HL}.}
\end{figure}

We describe briefly how to obtain it.
The distributional recursion $X=^{\!\!\!\!d}x^2/(x^2+\sum_{i=1}^K X_i)$ with $K\sim P=\textrm{Poisson(2)}$ is iterated a finite number $r$ of times with initial values $X_i\equiv 1$. The obtained random variable $X(r)$ represents the monomer density on a truncated Galton-Watson tree $\T(P,P,r)$ (lemma \ref{lemma: recurrence relations}).
If $X$ is the fixed point of the equation, we know that $X(2r)\searrow X$, $\,X(2r+1)\nearrow X$ as $r\to\infty$ (proposition \ref{prop: alternate monotonicity on tree}, theorem \ref{th: Galton-Watson}) and that $\E[X]$ is the asymptotic monomer density on $(G_n)_{n\in\N}$ (theorem \ref{main theorem}).\\
For values of $x=0.01,\,0.1,\,0.2,\,\dots,\,2$, the random variables $X(r),\,r=3,\,4,\,5,\,6$ are simulated numerically $10000$ times and an empirical mean is done in order to approximate $\E[X(r)]$.
The results are plotted as circles, squares, diamonds, triangles connected by straight lines.\\
The dot at $0.216074$ on the vertical axes corresponds to the exact value of the monomer density when the monomer activity $x\to0$, supplied by the Karp-Sipser formula \cite{KS} or by its extension due to Bordenave, Lelarge, Salez \cite{BLS}.
Therefore the graph of the monomer density $x\mapsto\E[X]=\displaystyle\lim_{n\to\infty}\varepsilon_{G_n}(x)$ starts from $(0,\,0.216074)$ and lays between the diamonds and triangles curves.

\section*{Appendix: general correlation inequalities on trees}
Consider the general monomer-dimer model (see remark \ref{rk: general m-d model}) on a finite graph $G=(V,E)$.
The Heilmann-Lieb recursion \cite{HL}, given a vertex $o$ and its neighbours $u$, reads
\begin{equation}\label{eq: H-L general recursion}
Z_G \,=\, x_o\,Z_{G-o} \,+\, \sum_{u\sim o}\,w_{ou}\,Z_{G-o-u}\;.
\end{equation}
Another simple and useful remark is that if $G$ is the \textit{disconnected} union of two subgraphs $G',\,G''$ then the partition function factorizes:
\begin{equation}\label{eq: factorisation}
G=G'\sqcup G'' \;\Rightarrow\; Z_G = Z_{G'}\,Z_{G''}\;.
\end{equation}
Now consider the probabilities of having a monomer on a given vertex $o$ and a dimer on a given edge $ou$ and denote them respectively
\[ \RR(G,o):=\langle\1_{o\in\MM_G(D)}\rangle_G\;,\qquad \EE(G,ou):=\langle\1_{ou\in D}\rangle_G\;.\]
Direct computations shows that these quantities can be expressed using first derivatives of the pressure:
\begin{equation}\label{eq: pressure first derivatives}
x_o\;\frac{\partial}{\partial x_o}\,\log Z_G\,=\,\RR(G,o)\;,\qquad w_{ou}\;\frac{\partial}{\partial w_{ou}}\,\log Z_G\,=\,\EE(G,ou)\;;
\end{equation}
while the second derivatives of the pressure are related to covariances:
\begin{equation}\begin{split}\label{eq: pressure second derivatives}
&x_p\;\frac{\partial}{\partial x_p}\,\RR(G,o) \,=\,
\langle\1_{o\in \MM(D)}\,\1_{p\in \MM(D)}\rangle_G \,-\, \langle\1_{o\in \MM(D)}\rangle_G\;\langle\1_{p\in \MM(D)}\rangle_G\\[2pt]
&w_{pv}\;\frac{\partial}{\partial w_{pv}}\,\EE(G,ou) \,=\,
\langle\1_{ov\in D}\,\1_{pv\in D}\rangle_G \,-\, \langle\1_{ou\in D}\rangle_G\;\langle\1_{pv\in D}\rangle_G\\[2pt]
&w_{pv}\;\frac{\partial}{\partial w_{pv}}\,\RR(G,o) \,=\,
x_o\;\frac{\partial}{\partial x_o}\,\EE(G,pv) \,=\\
&\qquad\qquad\qquad\qquad =\,\langle\1_{o\in \MM(D)}\,\1_{pv\in D}\rangle_G \,-\, \langle\1_{o\in \MM(D)}\rangle_G\;\langle\1_{pv\in D}\rangle_G
\end{split}\end{equation}
where $p$ is another vertex a $v$ is one of its neighbours.

Under the hypothesis that the underlying graph is a tree, it's possible to prove a family of general correlation inequalities for the monomer-dimer model: the direction of these inequalities depends on whether the graph distance between the considered edges and vertices is even or odd.
\begin{prop}\label{prop: general correlation inequalities on tree}
Suppose the graph $G=T$ is a tree.
Let $ou,\,pv\in E$. Then:
\[\frac{\partial}{\partial x_p}\,\RR(T,o)\,
\left\{\begin{array}{ll}
\geq 0, &\text{if }\,o=p\ \lor\ d_T(o,p)\text{ is odd}\\[5pt]
\leq 0, &\text{if }\,o\neq p\ \land\ d_T(o,p)\text{ is even}
\end{array}\right.\,;\]
\[\frac{\partial}{\partial w_{pv}}\,\EE(T,ou)\,
\left\{\begin{array}{ll}
\geq 0, &\text{if }\,ou=pv\ \lor\ d_T(ou,pv)\text{ is odd}\\[5pt]
\leq 0, &\text{if }\,ou\neq pv\ \land\ d_T(ou,pv)\text{ is even}
\end{array}\right.\,;\]
\[\frac{\partial}{\partial w_{pv}}\,\RR(T,o) \,=\, \frac{\partial}{\partial x_o}\,\EE(T,pv) \,
\left\{\begin{array}{ll}
\leq 0, &\text{if }\,d_T(o,pv)\text{ is even}\\[5pt]
\geq 0, &\text{if }\,d_T(o,pv)\text{ is odd}
\end{array}\right.\,;\]
where $d_T(o,p)$ denotes the distance between two vertices $o,p$ on $T$, that is the length (number of edges) of the shortest path on $T$ connecting them, while $d_T(o,pv)=\min\{d_T(o,p),\,d_T(o,v)\}\,$, $d_T(ou,pv)=\min\{d_T(o,pv),\,d_T(ou,p)\}$.
\end{prop}

Before the proof let us introduce some notations and a lemma.
Given a tree $T$ and two vertices $c_0,\,c_l$ such that $d_T(c_0,c_l)=l$, there exists a unique simple path on $T$ connecting them and we will denote it by
\[c_0,\,c_1,\,\dots,\,c_l\]
where each $c_sc_{s+1}$ is an edge of $T$ and the vertices $c_s$ are all distinct.
It will be useful to consider the rooted tree $(T,c_0)$. As usual this choice of a root induces an order relation on the vertex set of $T$: given two vertices $u,\,v$, the relation ``u is son of v'' will be shortened as $u\leftarrow v$ and the sub-tree of $T$ induced by $v$ and its descendants will be denoted $T_v$.

\begin{lem}\label{tree_fundamental}
The inequality
\begin{equation}\label{eq: tree_fundamental}
\1_{l\geq1}\;Z_{T_{c_1}-T_{c_l}}\;Z_{T} \;\gtreqless\; Z_{T_{c_1}}\,Z_{T-T_{c_l}}
\end{equation}
holds with the direction $\geq$ if $l$ is odd / $\leq$ if $l$ is even.
\end{lem}

\proof
If $l=0$, clearly the inequality (\ref{eq: tree_fundamental}) holds with $\leq\,$.\\
Now assume $l\geq1$. Rewrite the partition functions appearing in (\ref{eq: tree_fundamental}) making explicit all the different possibilities (monomer/dimer) that may interest the root $c_0$. To do it use formulae (\ref{eq: H-L general recursion}) and (\ref{eq: factorisation}):
\[\begin{split}
Z_T \,=&\,\; x_{c_0}\, \big(\!\!\prod_{\,v\leftarrow c_0\atop v\neq c_1}\!\!Z_{T_v}\big)\, Z_{T_{c_1}} \\
&+ \sum_{v\leftarrow c_0\atop v\neq c_1}w_{c_0v}\, \big(\!\!\prod_{\,v'\!\leftarrow v\atop}\!\!Z_{T_{v'}}\big)\, \big(\!\!\prod_{\,\tilde{v}\leftarrow c_0\atop \tilde{v}\neq v,c_1}\!\!Z_{T_{\tilde{v}}}\big)\, Z_{T_{c_1}} \\
&+\, w_{c_0c_1}\, \big(\!\!\prod_{\,v\leftarrow c_0\atop v\neq c_1}\!\!Z_{T_v}\big)\, \big(\!\!\prod_{\,u\leftarrow c_1\atop u\neq c_2}\!\!Z_{T_u}\big)\, Z_{T_{c_2}}
\end{split}\]
\[\begin{split}
Z_{T-T_{c_l}} =&\,\; x_{c_0}\, \big(\!\!\prod_{\,v\leftarrow c_{0}\atop v\neq c_1}\!\!Z_{T_v}\big)\, Z_{T_{c_1}-T_{c_l}} \\
&+ \sum_{v\leftarrow c_0\atop v\neq c_1} w_{c_0v}\, \big(\!\!\prod_{\,v'\!\leftarrow v\atop}\!\!Z_{T_{v'}}\big)\, \big(\!\!\prod_{\,\tilde{v}\leftarrow c_0\atop \tilde{v}\neq v,c_1}\!\!Z_{T_{\tilde{v}}}\big)\, Z_{T_{c_1}-T_{c_l}} \\
&+\, \1_{l\geq2}\;w_{c_0c_1}\, \big(\!\!\prod_{\,v\leftarrow c_0\atop v\neq c_1}\!\!Z_{T_v}\big)\, \big(\!\!\prod_{\,u\leftarrow c_1\atop u\neq c_2}\!\!Z_{T_u}\big)\, Z_{T_{c_2}-T_{c_l}}
\end{split}\]
Substituting these expressions into the inequality (\ref{eq: tree_fundamental}) and simplifying, it rewrites
\begin{equation}\label{eq: tree_fundamental1}
Z_{T_{c_2}}\,Z_{T_{c_1}-T_{c_l}} \;\gtreqless\; \1_{l\geq2}\;Z_{T_{c_2}-T_{c_l}}\;Z_{T_{c_1}}
\end{equation}
If $l=1$ clearly the inequality (\ref{eq: tree_fundamental1}), and therefore the inequality (\ref{eq: tree_fundamental}), holds with direction $\geq\,$.\\
Now assume $l\geq2$. Observe that the inequality (\ref{eq: tree_fundamental1}) has the same shape of (\ref{eq: tree_fundamental}), except that it has the opposite direction and the sub-tree $T_{c_1}$ is considered instead of the tree $T\equiv T_{c_0}$.\\
Therefore one iterates the argument $l+1$ times, obtaining that
\begin{itemize}
\item if $l$ is odd, then the inequality (\ref{eq: tree_fundamental}) is equivalent to the following
$$Z_{T_{c_{l+1}}}\,Z_{T_{c_l}-T_{c_l}} \;\gtreqless\; \1_{l\geq l+1}\;Z_{T_{c_{l+1}}-T_{c_l}}\;Z_{T_{c_l}}$$
which clearly holds with direction $\geq\,$;
\item if $l$ is even, then the inequality (\ref{eq: tree_fundamental}) is equivalent to the following
$$\1_{l\geq l+1}\;Z_{T_{c_{l+1}}-T_{c_l}}\;Z_{T_{c_l}} \;\gtreqless\; Z_{T_{c_{l+1}}}\,Z_{T_{c_l}-T_{c_l}}$$
which clearly holds with direction $\leq\,$.\qedhere
\end{itemize}
\endproof

We write the proof only for the third statement of the proposition: the first two can be proved with analogous arguments.
\proof [Proof of Proposition \ref{prop: general correlation inequalities on tree} (third statement)]
Assume without loss of generality that $d_T(pv,o)=d_T(p,o)=l$.
Set $c_0:=o,\,c_l:=p$ and consider the rooted tree $(T,o)$ with the notations previously introduced.
Using relations (\ref{eq: pressure second derivatives}) and (\ref{eq: H-L general recursion}) it's easy to compute
\[w_{pv}\;\frac{\partial}{\partial w_{pv}}\,\RR(T,o) \,=\, \1_{o\neq p}\;\frac{x_o\,w_{pv}\,Z_{T-o-p-v}}{Z_T} \,-\, \frac{x_o\,Z_{T-o}}{Z_T}\;\frac{w_{pv}Z_{T-p-v}}{Z_T}\;.\]
Therefore to determine the sign of $\partial\RR(T,o) / \partial w_{pv}$ it suffices to study the inequality%
\begin{equation}\label{eq: correlation_tree5}
\1_{o\neq p}\;Z_{T-o-p-v}\;Z_T \;\gtreqless\; Z_{T-o}\;Z_{T-p-v}\;.
\end{equation}
The connected components of each graph appearing in this inequality are:
\begin{gather*}
T-o-p-v \,= \bigsqcup_{u\leftarrow o\atop u\neq c_1} T_u \,\sqcup\, (T_{c_1}-T_p) \,\sqcup \bigsqcup_{a\leftarrow p\atop a\neq v}T_a \,\sqcup \bigsqcup_{b\leftarrow v\atop}T_b \\[4pt]
T-o \,= \bigsqcup_{u\leftarrow o\atop u\neq c_1} T_u \,\sqcup\, T_{c_1} \\[4pt]
T-p-v \,=\, (T-T_p) \,\sqcup \bigsqcup_{a\leftarrow p\atop a\neq v} T_a \,\sqcup \bigsqcup_{b\leftarrow v\atop} T_b
\end{gather*}
Hence, applying (\ref{eq: factorisation}) and simplifying, the inequality (\ref{eq: correlation_tree5}) rewrites
\begin{equation}\label{eq: correlation_tree6}
\1_{l\geq1}\;Z_{T_{c_1}-T_p}\;Z_{T} \;\gtreqless\; Z_{T_{c_1}}\;Z_{T-T_p}\;.
\end{equation}
This inequality is of the same kind of (\ref{eq: tree_fundamental}), therefore conclude by lemma \ref{tree_fundamental}.
\endproof

{\bf Acknowledgements} The authors thank Sander Dommers for many valuable discussions.

\end{document}